\documentclass[aps,nofootinbib, prd, amsmath,floats, twocolumn,floatfix,superscriptaddress, showpacs,reprint]{revtex4}

\usepackage{graphicx}
\usepackage{amsmath,amssymb}
\usepackage{amsfonts}
\usepackage{xspace} 
\usepackage[usenames]{color}
\usepackage{dcolumn}
\usepackage{bm}
\usepackage{xcolor}
\usepackage{mathrsfs}
\usepackage[colorlinks=true]{hyperref}
\usepackage[all]{hypcap} 
 \usepackage[utf8]{inputenc}

\def\be{\begin{equation}}
\def\ee{\end{equation}}
\def\beq{\begin{eqnarray}}
\def\eeq{\end{eqnarray}}

\newcommand{\hp}{h_{+}}
\newcommand{\hx}{h_\times}
\newcommand{\oml}{\omega_{lmn}}

\newcommand{\tl}{\tau_{lmn}}
\newcommand{\phl}{\phi_{lmn}}

\newcommand{\tn}{\textnormal}

\begin{document}

\title{Observing binary black hole ringdowns by advanced gravitational wave detectors}

\author{Andrea Maselli}
\email{andrea.maselli@uni-tuebingen.de}
\affiliation{Theoretical Astrophysics, IAAT, University of T\"uebingen, 
T\"uebingen 72076, Germany}

\author{Kostas D. Kokkotas}
\email{kostas.kokkotas@uni-tuebingen.de}
\affiliation{Theoretical Astrophysics, IAAT, University of T\"uebingen, 
T\"uebingen 72076, Germany}

\author{Pablo Laguna}
\email{plaguna@gatech.edu}
\affiliation{Center for Relativistic Astrophysics and School of Physics,
Georgia Institute of Technology, Atlanta, Georgia 30332, USA}

\date{\today}


\begin{abstract}
The direct discovery of gravitational waves from compact binary systems leads for the first time 
to explore the possibility of black hole spectroscopy. Newly formed black holes produced by coalescing 
events are copious emitters of gravitational radiation, in the form of damped sinusoids, the quasi normal 
modes. The latter provide a precious source of information on the nature of gravity in the strong field 
regime, as they represent a powerful tool to investigate the validity of the no-hair theorem. 
In this work we perform a systematic study on the accuracy with which current and future interferometers will 
measure the fundamental parameters of ringdown events, such as frequencies and damping times. 
We analyze how these errors affect the estimate of the mass and the angular momentum of the 
final black hole, constraining the parameter space which will lead to the most precise measurements. 
We explore both single and multimode events, showing how the uncertainties evolve when multiple 
detectors are available. We also prove that, for second generation of interferometers, 
a network of instruments is a crucial and necessary ingredient to perform 
strong-gravity tests of the no-hair theorem. Finally, we analyze the constraints that a third generation of 
detectors may be able to set on the mode's parameters, comparing the projected bounds  against those 
obtained for current facilities. 

\end{abstract}

\maketitle

\section{Introduction}

Black holes (BH) represent the most clear, macroscopic expression of one of the fundamental forces 
of Nature, and they provide a unique testbed to investigate gravity in a pure strong-field regime.
The discovery of a $67M_\odot$ and a $22M_\odot$ binary coalescence has directly proven for 
the first time that BH may form in the Universe, and merge within the Hubble time 
\cite{Abbott:2016blz,Abbott:2016nmj}. Most exciting, these detections allow to realize  a field of 
research, the {\it BH spectroscopy}, considered pure theoretical so far \cite{Kokkotas:1999bd}. 

Astrophysical BHs outside of equilibrium are characterized by a spectrum of damped oscillations, 
called quasi normal modes (QNMs) \cite{Vishveshwara:1970zz,Teukolsky:1973ha,Chandrasekhar:1975zza,Price:1994pm}. The output of compact binary coalescence, is one of the best examples of this phenomenon. 
After the merger phase, a newly born and highly distorted BH emits gravitational radiation as a 
superposition of damped sinusoids, which decay exponentially, until the object reaches its quiet, 
stationary state. QNMs in General Relativity (GR) are uniquely determined by the BH mass $M$ and 
intrinsic angular momentum $J=jM^2$ (being $j$ the spin parameter). This feature is a direct 
consequence of the so called no-hair theorem, which states that rotating compact object belongs to 
the Kerr family, whose exterior stationary and isolated gravitational field depends only on two parameters, 
$M$ and $J$ \cite{wiltshire2009kerr}.
We note that BHs in real physical environments will not be perfectly stationary or in a clear vacuum state: 
other objects or fields such as stars or accretion disks may alter 
their Kerr nature. However, under the assumption that such perturbations are small enough to  be practically 
unobservable, we can safely assume that such objects are indeed described by the Kerr metric \cite{Cardoso:2016ryw}.
In this sense, QNMs play a key role, as their detection would confirm the nature of the source, and at the 
same time would help to assess the validity of the Kerr hypothesis. 

Although the existence of BHs has been now experimentally settled down, a proof of the validity 
of the no-hair theorem is still lacking. Coalescing compact binaries offer a new window on this picture. 
It is crucial to note that precise mass/spin measurements of ringing BHs would also be a precious 
source of information for two other GR cornerstones: (i) the Hawking area theorem 
\cite{PhysRevLett.26.1344} for which the final area should be larger then the sum of those of 
the progenitors, (ii) the Penrose cosmic censorship which establishes that the BH spin parameter never 
exceeds the bound $j \leq1$.

QNMs have been extensively studied in literature. In particular, several efforts have been devoted to 
analyze the outcome of future detections by space interferometers as LISA \cite{AmaroSeoane:2012km,AmaroSeoane:2012je}, 
were more massive compact binaries are expected to produce louder events. 
It is worth to cite here the seminal papers by Echeverria \cite{Echeverria:1989hg}, Flanagan and Hughes 
\cite{Flanagan:1997sx,PhysRevD.57.4566} where a detailed discussion of the statistical techniques needed 
to extract the fundamental parameters of the waveforms has been developed.
A key reference for QNMs is given by the detailed investigation made by Berti and collaborators, 
in which the authors analyze both single and multimode events \cite{PhysRevD.73.064030,Berti:2007zu}. 
The latter represents the golden goal of BH spectroscopy, as at least two QNMs are required to make 
consistency tests of the no-hair theorem. The prospects of measuring the dominant ($l=m=2$) 
and a secondary component (as the $l=m=3$) have been tackled with different methods and 
data analysis strategies, including population synthesis simulations, focusing 
in particular on the signal to noise ratio necessary to distinguish the QNM features 
\cite{Dreyer:2003bv,Gossan:2011ha,Kamaretsos:2011um,Nakano:2015uja,Bhagwat:2016ntk}.
More recently, it has been pointed out that a coherent stack of several gravitational wave (GW)
signals increases the ringdown SNR, and therefore improves our ability to extract multiple QNMs 
from the interferometer's background noise \cite{Yang:2017zxs}.

A  general analysis concerning the connection between BH oscillations and  the formation of 
an event horizon has also been developed in \cite{Cardoso:2016rao,Konoplya:2016hmd}, showing 
that more exotic objects as wormholes can mimic BH ringing, both for the early and the late time modes. 
An interesting possibility is given by the analysis presented in \cite{Chirenti:2016hzd}, in which 
the authors have developed a general strategy to constrain such alternative scenarios 
investigating the QNMs spectrum. A first application of this model seems already to rule out the possibility 
that the first gravitational wave event GW150914 \cite{Abbott:2016blz} led to the formation of a rotating 
gravastar.
 
The aim of this paper is to systematically analyze  the ability of current 
and near-future terrestrial interferometers to observe QNM signals from stellar BHs. 
In particular, we compute the errors on the ringdown parameters, i.e. 
frequencies and damping time, focusing on how they affect our ability to measure the BH 
mass and spin angular momentum. We determine the configurations which favor possibly 
high accuracy observations, both for single and multi mode events. 

While the Advanced LIGO sites \cite{0264-9381-32-11-115012} are already under science mode, 
two other interferometers, Virgo \cite{0264-9381-32-2-024001} and Kagra \cite{0264-9381-29-12-124007}, 
are nearly to be completed. Multiple, independent observations, are a crucial ingredient for 
gravitational wave astronomy, as they increase the detection confidence and the overall signal to noise ratio. 
In this scenario, we investigate how our results change when a network of four instruments 
is taken into account.
Moreover, the first direct discovery of a GW signal is leading 
the quest to develop a third generation of interferometers, which will improve the existing 
sensitivity by more than an order of magnitude \cite{LIGOWhite}. It is therefore timely 
to compare the performance of such detectors with the current facilities.

This paper is organized as follows. In Sec.~\ref{Sec:QNM} we introduce the basic formalism to 
describe QNMs in GR, and the data-analysis tools used. In Sec.~\ref{Sec:results} 
we present our numerical results, analyzing the errors on the BH physical parameters obtained 
from ringdown events for second and third generation of interferometers, respectively. 
In Sec.~\ref{Sec:conc} we summarize our findings. 
Throughout the paper we will use geometrical units $(G=c=1)$.

\section{The mathematical toolkit}
In this section, we shall summarize the basic properties of BH ringing, introducing the  
concepts needed for our numerical analysis. We refer the reader to the review article 
\cite{Kokkotas:1999bd,Berti:2009kk} (and references therein) for an extensive lecture on the topic. 

\subsection{Quasi Normal Modes formalism}\label{Sec:QNM}

The gravitational wave signal measured by the interferometer consists of a linear superposition of the two 
polarization states $\hx$ and $\hp$:
\begin{equation}\label{totalh}
h=\hp F_+ +\hx F_\times\ , 
\end{equation}
where $F_{+,\times}$ are pattern functions which depend on the source orientation with 
respect to the detector \cite{Sathyaprakash:2009xs}.
For a given mode, specified by the numbers $(l,m)$ and the overtone index $n$ 
we have:
\begin{align}
\hp&=\frac{M}{d}\Re\left[{\cal A}_{lmn}^+e^{i(\oml t+\phl^+)}e^{-t/\tl}S_{lmn}\right]\label{hp}\ ,\\
\hx&=\frac{M}{d}\Im\left[{\cal A}_{lmn}^\times e^{i(\oml t+\phl^\times)}e^{-t/\tl}S_{lmn}\right]\label{hc}\ ,
\end{align}
where $\oml=2\pi f_{lmn}$ and $\tau_{lmn}$ are the mode's frequency and damping time,  
the amplitudes ${\cal A}^{+,\times}_{lmn}$ and the phase coefficients $\phl^{+,\times}$ are 
real quantities, while $d$ is the luminosity distance of the source.
The 2-spin-weighted spheroidal harmonics $S_{lmn}=e^{i m\alpha}\bar{S}_{ln}(\beta)$ 
are functions of the azimuthal and the polar angles $(\alpha,\beta)$. These have been proved to 
form an  orthonormal set of eigenfunctions, i.e.
\begin{equation}\label{Scond}
\int S_{lm}(\alpha,\beta)S^\star_{l'm'}(\alpha,\beta)d\Omega =\delta_{l l'}\delta_{mm'} \quad (n=n')\ ,
\end{equation}
where the symbol $\star$ denotes complex conjugation \cite{Berti:2005gp}. 
Following \cite{Flanagan:1997sx,PhysRevD.73.064030} we assume that for $t<0$ the waveform is 
identical to that computed for $t>0$. Therefore we replace the damping factor $e^{-t/\tl}$ 
with $e^{-\vert t\vert/\tl}$, dividing by a factor $\sqrt{2}$ in the amplitude to compensate the doubling. 
As noted in \cite{Flanagan:1997sx}, this procedure allows to fully  characterize the QNM spectrum, 
without any prescription for the unknown merger-waveform before the ringdown.
We also average 
over the detector and the BH directions, making use of the following identities
\begin{align}
\langle F_+^2\rangle=\langle F_\times^2\rangle=\frac{1}{5}\ ,\ \langle F_+F_\times\rangle=0\ \ ,\ 
\langle \vert S_{lmn}\vert^2 \rangle=\frac{1}{4\pi}\ .
\end{align}
Finally, we introduce the rescaled amplitudes 
\begin{equation}
A^{+}_{lmn}=\frac{M}{d}{\cal A}^{+}_{lmn}\quad \ , \quad A^{\times}_{lmn}=\frac{M}{d}{\cal A}^{\times}_{lmn}
=N_\times A^+_{lmn}\ ,
\end{equation}
being $N_\times$ a relative scale factor, and the phase-shift $\phi_0$:
\begin{equation}
\phi^{\times}_{lmn}=\phi^{+}_{lmn}+\phi_0\ .
\end{equation}
Assuming that $(N_\times,\phi_0)$ are known, the waveform (for a single mode) depends 
on four quantities $(A^{+}_{lmn},\phi^{+}_{lmn},f_{lmn},\tau_{lmn})$.
Eqns.~\eqref{totalh} and \eqref{hp}-\eqref{hc} represent the basic ingredients to build all the 
physical quantities needed for our analysis, as the SNR
\begin{equation}\label{SNR1}
\rho^2=4\int_0^\infty \frac{\tilde{h}(f)\tilde{h}^\star(f)}{S_n(f)}df\ ,
\end{equation}
where $S_n(f)$ is the noise spectral density of the interferometer, and $\tilde{h}(f)$ is the Fourier 
transform of the time-dependent waveform. We note that in this case, the change to the frequency 
domain is rather straightforward:
\begin{widetext}
\begin{align}
\tilde{h}_+(f)&=\frac{A^+_{lmn}}{\sqrt{2}}\left[e^{i\phi^+_{lmn}}S_{lmn}(\alpha,\beta)b_+(f)
+e^{-i\phi^+_{lmn}}S^\star_{lmn}(\alpha,\beta)b_-(f)\right]\ ,\label{hpf}\\
\tilde{h}_\times(f)&=-\frac{i}{\sqrt{2}}N^+A^+_{lmn}\left[e^{i(\phi^+_{lmn}+\phi_0)}S_{lmn}(\alpha,\beta)b_+(f)
+e^{-i(\phi^+_{lmn}+\phi_0)}S^\star_{lmn}(\alpha,\beta)b_-(f)\right]\ ,\label{hcf}
\end{align}
\end{widetext}
where $b_\pm(f)$ are the Breit-Wigner functions: 
\begin{equation}\label{BreitWigner}
b_\pm(f)=\frac{\tau_{lmn}}{1+(2\pi \tau_{lmn})^2(f\pm f_{lmn})^2}\ .
\end{equation}
Replacing the former into Eq.~\eqref{SNR1}, and averaging over the angles we obtain:
\begin{equation}\label{SNR2}
\rho^2=\frac{1}{10\pi}A^{+^2}_{lmn}\int_0^\infty\frac{df}{S_n(f)}(b_+^2+b_-^2)(1+N_\times^2)\ .
\end{equation}
As noted in \cite{PhysRevD.57.4566}, the SNR can be recast in terms of the gravitational wave 
energy spectrum $dE/df$:
\begin{equation}\label{SNR3}
\rho^2=\frac{2}{5\pi^2 d^2}\int^\infty_0\frac{1}{f^2S_n(f)}\frac{dE}{df}df\ .
\end{equation}
which is related to the radiation efficiency $\epsilon_\tn{QNM}$:
\begin{equation}\label{erd}
\epsilon_\tn{QNM}=\frac{1}{M}\int_0^\infty \frac{dE}{df} df\ ,
\end{equation}
which controls the amount of energy released in QMNs.
Combining Eqns.~\eqref{SNR2}-\eqref{erd} we have 
\begin{equation}\label{eps2}
\epsilon_\tn{QNM}=\frac{\pi M}{4}(1+N_\times^2){\cal A}^{+^2}_{lmn}\int_0^\infty f ^2(b_+^2+b_-^2) df\ .
\end{equation}
For any choice of the source parameters $(M,d,f_{lmm},\tau_{lmn},\phi^+_{lmn},\phi_0,N_\times)$, 
and a given efficiency $\epsilon_\tn{QNM}$, we can numerically solve Eq.~\eqref{eps2} to find the signal 
amplitude ${\cal A}^+_{lmn}$.
 
\subsection{Parameter estimation}
For a generic signal of the form $h(t,\vec{x})$, we want to determine the source parameters 
$\vec{x}=\{x_1,\ldots,x_n\}$, and the measurement errors $\Delta\vec{x}=\vec{x} -\vec{y}$, 
were $\vec{y}$ are assumed to be the true values. To
this aim we need to compute the probability $p(\vec{x}\vert s)$, given the detector
output $s(t)=h(t,\vec{x})+n(t)$, with $n(t)$ instrumental noise, which we consider to be stationary. 
Up to a normalization constant (independent of the physical parameters), the conditional 
probability can be written as:
\begin{equation}\label{prob}
p(\vec{x}\vert s)\propto p^{(0)}(\vec{x})e^{-\frac{1}{2}(h(\vec{x})
-s\vert h(\vec{x})-s)}\ ,
\end{equation}
where $p^{(0)}(\vec{x})$ represents the prior on the source parameters \cite{PhysRevD.49.2658}, 
and the inner product on the waveform space $(\cdot \vert \cdot)$ reads:
\begin{equation}\label{inner}
(g\vert h)=2\int_{-\infty}^{\infty}\frac{\tilde h(f)\tilde g^{\star}(f)+\tilde h^{\star}(f)\tilde g(f)}{S_{n}(f)}df\ .
\end{equation} 
Following  to the principle of the maximum-likelihood, the source parameters are represented 
by those which maximize the probability distribution \eqref{prob}. Moreover, for 
gravitational wave signals characterized by large SNR, we expect $p(\vec{x}\vert h)$ to be 
narrowly peaked around the true values $\vec{y}$. Therefore, we can expand Eq.~\eqref{prob} 
as Taylor series up to the second order in $\Delta \vec{x}=\vec{x} -\vec{y}$, such that 
\begin{equation}\label{prob1}
p( \vec{x}\vert s)\propto p^{(0)}(\vec{x})e^{-\frac{1}{2}\Gamma_{ab}\Delta x^{a}\Delta x^{b}}\ ,
\end{equation}
where 
\begin{equation}\label{FisherM}
\Gamma_{ab}=\left(\frac{\partial h}{\partial x^a}\bigg\vert \frac{\partial h}{\partial x^b}\right)
\end{equation}
evaluated at $\vec{x}=\vec{y}$, is the Fisher  information matrix. The latter, is directly 
related to the covariance matrix $\Sigma^{ab}$:
\begin{equation}
\Sigma^{ab}=\left(\Gamma^{-1}\right)^{ab}\ ,
\end{equation}
where $\Gamma^{-1}$ is the inverse of Fisher matrix. In this way we
can define the error associated to the parameter $x^{a}$ as
\begin{equation}\label{error}
\sigma_{a}=\sqrt{\Sigma^{aa}}\ ,
\end{equation}
and the correlation coefficient between $x^{a}$ and $x^{b}$ as
\begin{equation}\label{corr}
c_{ab}=\frac{\langle \Delta x^{a}\Delta x^{b}\rangle}{\Sigma^{aa}\Sigma^{bb}}=
\frac{\Sigma^{ab}}{\sqrt{\Sigma^{aa}\Sigma^{bb}}}\ ,
\end{equation}
with $c_{ab}\in[-1,1]$.  Eq.~(\ref{prob1}) holds whether
$p^{(0)}(\vec{x})$ is uniform around $\vec{y}$ or not.  In the special
case when $p^{(0)}$ is Gaussian:
\begin{equation}
p^{(0)}(\vec{x})\propto e^{-\frac{1}{2}\Gamma_{ab}^{(0)}(x^{a}-\bar x^{b})(x^{b}-\bar x^{b})}\ ,
\end{equation}
the probability distribution $p(\vec{x}\vert s)$ is Gaussian with total 
covariance matrix given by:
\begin{equation}
\Sigma_{ab}=(\Gamma_{ab}+\Gamma_{ab}^{(0)})^{-1} \ .
\end{equation}

As seen in the previous section the waveform for BH ringing depends on four parameters. 
Rather than the damping time $\tau_{lmn}$, we will compute the Fisher matrix in terms of the 
quality factor $Q_{lmn}=\pi f_{lmn}\tau_{lmn}$. This yields a $4\times 4$ matrix in the 
variables $\vec{x}=(\ln A^{+}_{lmn},\phi^{+}_{lmn},f_{lmn},Q_{lmn})$\footnote{In computing the 
Fisher matrix, we have neglected the derivative of the spheroidal functions $S_{lmn}$ with respect 
to $f_{lm}$ and $Q_{lm}$. However, as noted in \cite{PhysRevD.73.064030}, these terms are 
at least quadratic in the spin $j$ parameter, and may be considered small.}. 

Moreover, as shown in \cite{PhysRevD.73.064030}, $f_{lmn}$ and $Q_{lmn}$ can be expressed in 
terms of the BH mass and spin variable $M,j$ through numerical fits of the form 
\begin{equation}\label{fitfQ}
f_{lmn}=\frac{f_1+f_2(1-j)^{f_3}}{2\pi M}\ \ ,\ Q_{lmn}=q_1+q_2(1-j)^{q_3}\ ,
\end{equation}
where $(q_i,f_i)$ are fitting coefficients which depends on the particular set of number 
$(l,m,n)$ (see Table VIII-X of \cite{PhysRevD.73.064030}). Therefore, we will also compute 
the Fisher matrix in terms of the parameters $\vec{z}=(\ln A^{+}_{lmn},\phi^{+}_{lmn},M,j)$. 
The transformation between the two sets of coordinates is simply given by 
\begin{equation}\label{changebasis}
\Gamma(\vec{z})_{ab}=[C^\tn{T}]_{ac}[ \Gamma(\vec{x})]_{cd}[C]_{db}\ , 
\end{equation}
where $C$ is the change of basis matrix, with elements $C_{ij}=\frac{\partial x_i}{\partial z_j}$, 
and $C^\tn{T}$ its transpose. 

Finally, it is useful to consider how to combine the information on the source's parameters, when 
multiple detectors are taken into account. This is indeed the case of gravitational wave interferometers, 
as, beside the two LIGO observatories, Virgo and KAGRA are in the final completion phase.
If the experiments are all independent, the total probability distribution will be given by the 
product of multiple $p(\vec{x}\vert s)$ computed for each detector, and the individual 
Fisher matrices will just add linearly 
\begin{equation}
\Gamma_{ab}=\Gamma_{ab}^\tn{(LIGO)}+\Gamma^\tn{(Virgo)}_{ab}+\cdots\ ,
\end{equation}
with the final covariance matrix obtained inverting $\Gamma_{ab}$. In the same 
spirit, the total SNR will be given by the sum in quadrature of the single-interferometer 
quantities, namely
\begin{equation}
\rho^2=\rho_\tn{(LIGO)}^2+\rho^2_\tn{(Virgo)}+\cdots\ .
\end{equation}

\subsection{Multiple modes}

BH spectroscopy may provide a crucial assessment of the no-hair theorem, confirming 
the uniqueness of the Kerr hypothesis for isolated astrophysical BHs in General Relativity. 
However, a genuine test of the such principle requires the detection of at least two 
QNMs, and this leads to identify how much energy is stored in the secondary oscillations, 
and what is the SNR required to distinguish the dominant component from the weaker one. 
It is worth to remark that it has been recently shown that, according to population synthesis studies, 
it seems unlike to have BH evolutionary scenarios leading to GW signals strong enough  
to distinguish between multiple QNMs with one single detector \cite{Berti:2016lat}. 
The numerical results for the multimode analysis presented here are therefore more speculative, and may be considered as 
reference and complementary to the analysis developed in \cite{Berti:2016lat}.
Nonetheless, we will prove how joint observations of multiple interferometers may lead 
to a more favorable prospect for testing GR in the strong field regime.

The multi-component analysis can be described starting from the discussion presented in Sec.~\ref{Sec:QNM}, as we replace 
the single mode polarizations \eqref{hp}-\eqref{hc}, with a infinite sum over all the possible 
configurations specified by the triplets $(l,m,n)$:
\begin{align}
\hp&=\sum_{lmn}\frac{M}{d}\Re\left[{\cal A}_{lmn}^+e^{i(\oml t+\phl^+)}e^{-t/\tl}S_{lmn}\right]\label{hp2}\ ,\\
\hx&=\sum_{lmn}\frac{M}{d}\Im\left[{\cal A}_{lmn}^\times e^{i(\oml t+\phl^\times)}e^{-t/\tl}S_{lmn}\right]\label{hc2}\ .
\end{align}
In this work we will consider only the case in which two modes with the same overtone number $n$ are excited simultaneously. 
In this scenario the Fisher matrix \eqref{FisherM} becomes a $8\times8$ matrix in the variables 
$(\ln A^{+}_{lmn},\phi^{+}_{lmn},f_{lmn},Q_{lmn},\ln  A^{+}_{l'm'n'},\phi^{+}_{l'm'n'},f_{l'm'n'},Q_{l'm'n'})$. 
However, as far as we consider components with the same overtone number $n=n'$, the angular average and 
the orthogonality of spheroidal functions (cfr. Eq.~\eqref{Scond}) decouple QNMs with different 
$l$ and $m$. This property simplifies our calculations since the total Fisher matrix 
$\Gamma_{ab}$ can be written as the sum of the matrices $\Gamma^{(1,2)}_{ab}$ 
computed for each mode :
\begin{equation}
\Gamma_{ab}=
\begin{pmatrix}
\Gamma^{(1)}_{ab} & 0 \\
0 & \Gamma^{(2)}_{ab}
\end{pmatrix}
\end{equation} 
This also applies to the total SNR, which in turn is defined as the sum 
$\rho=\sqrt{\rho^{(1)}+\rho^{(2)}}$.

\section{Numerical results} \label{Sec:results}
\begin{figure}[ht]
\centering
\includegraphics[width=7.5cm]{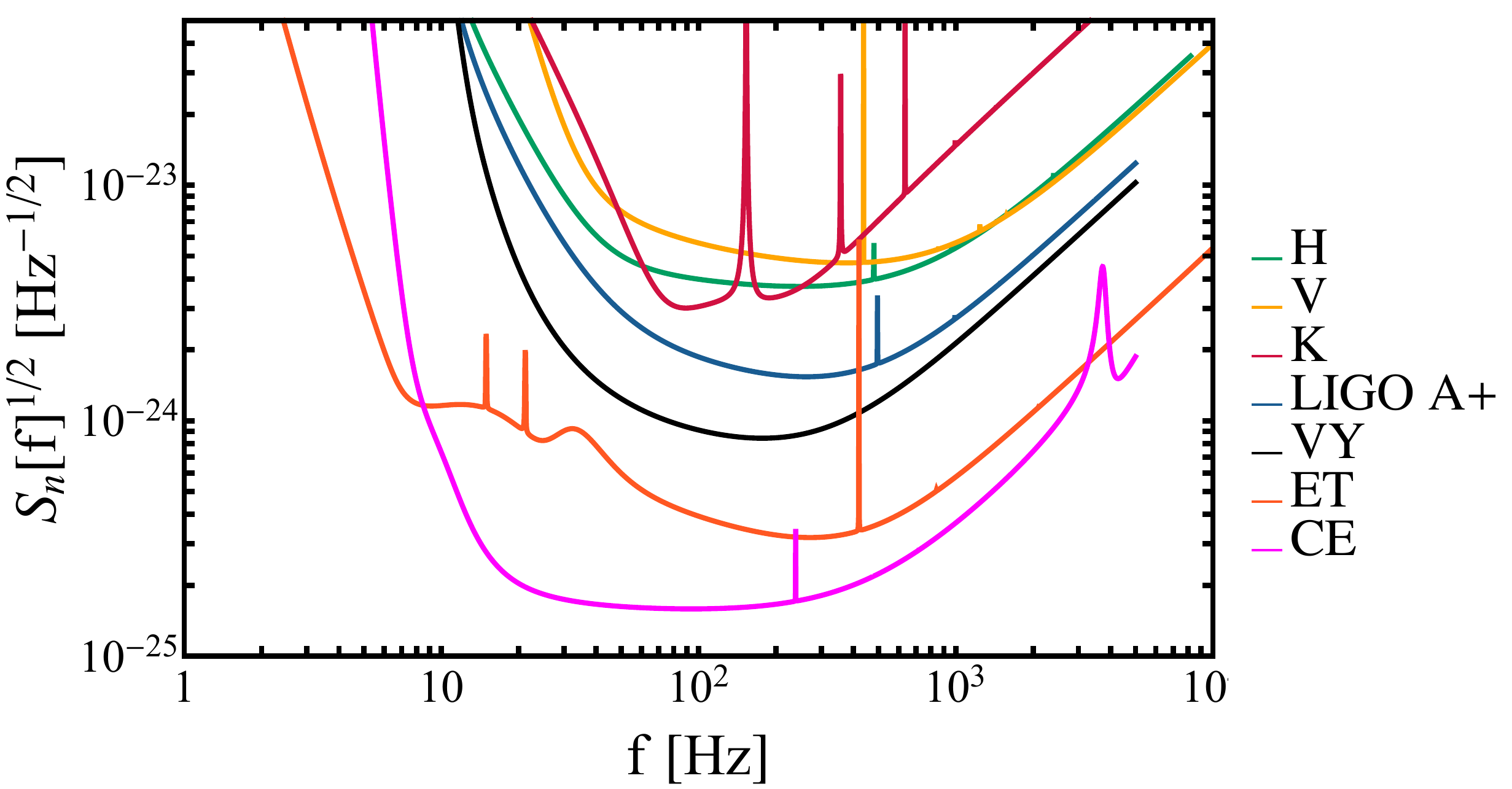}
\caption{Noise spectral densities of the terrestrial interferometers considered 
in this paper. From top to bottom we show:  AdLIGO, AdVirgo, and Kagra at design 
sensitivity, LIGO A+, Voyager, the Einstein Telescope and the Cosmic Explorer.}
\label{fig:PSD}
\end{figure}

Our results are based on the numerical integration of Eq.~\eqref{SNR2}, and \eqref{FisherM}. 
We will consider QNMs produced by BHs with mass $M\in[10,70]M_\odot$ and spin 
parameter $j\in[0.1,0.9]$ at prototype distances of $400$Mpc. We note that  the gravitational wave 
amplitude scales as $d^{-1}$, and therefore $\Sigma_{ab}=(\Gamma_{ab})^{-1}\propto d^{2}$. 
As a consequence the errors on the source parameters are proportional to the source distance 
$d$. In the single QNM analysis we compute the Fisher matrix for the fundamental mode with $n=0$, 
$l=m=2$, assuming a favorable scenario with radiation efficiency $\epsilon_\tn{QNM}=0.03$, and a 
negative scenario with $\epsilon_\tn{QNM}=0.01$. These values are consistent with the numerical 
results of \cite{Berti:2007fi}, where the ringdown efficiency was found to be proportional 
to the symmetric mass ratio of the (non spinning) binary progenitor, i.e. $\epsilon_\tn{QNM} =0.44 \nu^2$, 
with $\nu=m_1m_2/(m_1+m_2)^2$. In particular the optimistic and the pessimistic case can be related respectively 
to an equal mass system ($\nu=0.25$), and an asymmetric binary with $\nu\simeq0.1$.
Moreover, following \cite{PhysRevD.73.064030}, 
we fix $N_\times=1$ and $\phi^+_{lmn}=\phi_0=0$, both for single, and two-mode analysis. 
In the latter, we will focus on the subdominant components with $n=0$, $l=m=3$ and $l=m=4$, assuming 
astrophysical events in which the energy stored in the secondary mode is $1/10$ of the 
fundamental one\footnote{Since fixed, hereafter we will omit the subscript $n$ for the frequency mode 
$f_{lm}$ and quality factor $Q_{lm}$.}. We consider second and third generation of interferometers,  
taking into account for the former the two LIGO (H and L), Virgo (V) and KAGRA (K). For both the AdLIGO sites we assume the 
noise spectral density given by the \texttt{ZERO\_DET\_high\_P} anticipated design sensitivity curve 
\cite{zerodet}, while for Virgo and the Japanese detector we use the numerical data obtained in \cite{virgo} and  
\cite{kagra}, respectively. As far as future detectors are considered, we develop our analysis 
for the Einstein Telescope (ET) in {\it xylophone} mode \cite{Hild:2009ns}, AdLIGO with squeezing 
(LIGO A+) \cite{Miller:2014kma}, a LIGO-Voyager class mission (VY) \cite{LIGOWhite}, and the Cosmic Explorer (CE) 
with a 40-km wide-band configuration \cite{0264-9381-34-4-044001}. The sensitivity curves of all the 
detectors are shown for comparison in Fig.~\ref{fig:PSD}.

\subsection{Second generation detectors: single mode detection}
As far as only one interferometer is considered, we will assume that the signal is  
observed by a single AdLIGO detector. Moreover, to better quantify the measurements accuracy 
we will show our results in terms of the relative (percentage) error 
$\Delta i=\sigma_i/i$, where $i=( f_{lmn},Q_{fmn},M,j)$.

\begin{figure}[ht]
\centering
\includegraphics[width=4.25cm]{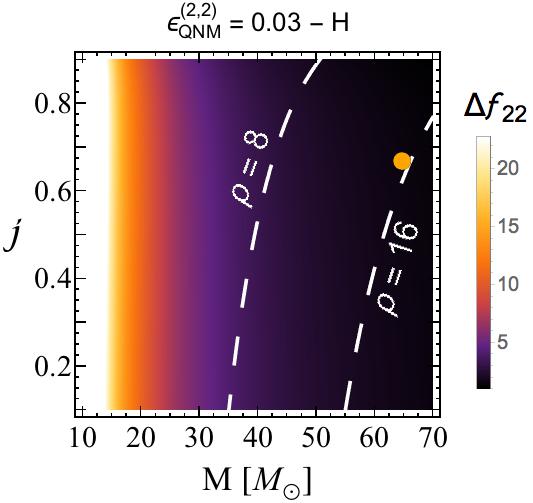}
\includegraphics[width=4.25cm]{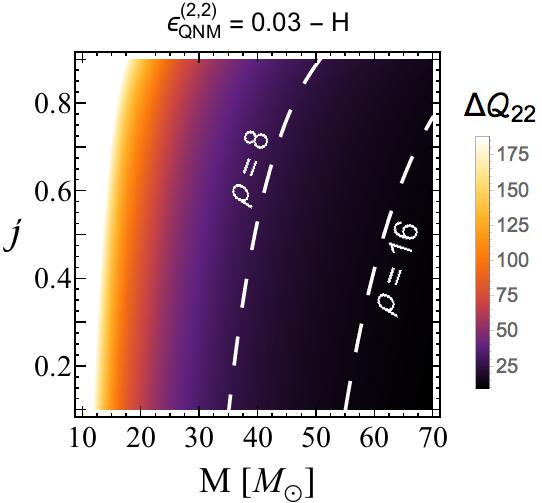}
\caption{Contour plots for $\Delta f_{22}$  in the $M-j$ plane for the frequency (left panel)
and the quality factor (right panel) of the fundamental (2,2) mode. The plots refer to 
a BH at $d=400$ Mpc, with radiation efficiency $\epsilon_\tn{QNM}=\epsilon^{(2,2)}_\tn{QNM}=0.03$. 
The white dashed curves identify lines of constant SNR, computed assuming a single 
detector with the AdLIGO sensitivity. The orange dot identifies a BH configuration with the same 
mass and spin of GW150914.}
\label{fig:fQ22}
\end{figure}

In Fig.~\ref{fig:fQ22} we show contour plots of $\Delta f_{22}$ and $\Delta Q_{22}$ as a function 
of the BH mass and spin, for the optimistic case $\epsilon_\tn{QNM}=0.03$. Moreover, we plot 
curves of constant SNR, given by the white dashed lines, with $\rho=8$ and $\rho=16$. 
For all the configurations considered we found relative errors on the fundamental frequency 
between $\sim1\%$ ad $20\%$, while for the quality factor the accuracy strongly decreases, 
leading to $100\%\lesssim\Delta Q_{22}\lesssim 20\%$. As expected, in both cases heavy 
systems, with $M\gtrsim 50M_\odot$, provide the most precise measurements, as the total 
SNR increases with the BH mass. We also note that these results seem rather insensitive 
to the intrinsic angular momentum $j$.

\begin{figure}[ht]
\centering
\includegraphics[width=4.25cm]{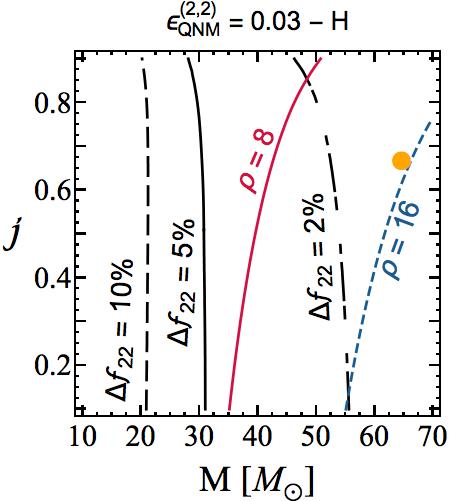}
\includegraphics[width=4.25cm]{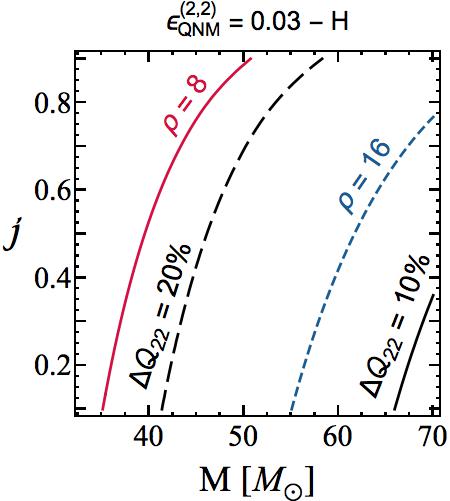}
\caption{Same as Fig.~\ref{fig:fQ22}, but for contour lines of fixed relative uncertaintiy 
for the frequency (left) and the quality factor (right) of the dominant mode with 
$l=m=2$.}
\label{fig:fQ22v2}
\end{figure}

To be more qualitative, in Fig.~\ref{fig:fQ22v2} we plot contour lines of fixed relative uncertainty 
$\Delta f_{22}=(10,5,2)\%$ and $\Delta Q_{22}(20,10)\%$. The left panel of the figure shows that a parameter 
space does exist, which allows the mode frequency $f_{22}$ to be determined with  good precision. In particular, 
for $M\lesssim 30M_\odot$ we do not expect  errors smaller than $5\%$, regardless the final BH spin. However, 
more massive objects lead in principle to very accurate measurement of $f_{22}$. Detections 
with relative accuracy of the order of 1-2\% require signals with $30M_\odot\lesssim M\lesssim 60M_\odot$
which correspond to an SNR range $\rho \in 8 \div 16$. Again, these results are nearly 
independent from the BH final spin. We note that a GW150914-like event, with $M=67.4M_\odot$ and $j=0.67$ 
\cite{TheLIGOScientific:2016wfe} would lie at the edge of the $\rho=16$ curve (orange dot in the 
figure), leading to $\Delta f_{22}\sim1.4\%$\footnote{Using the real distance measured by the LIGO 
collaboration $d=410$ Mpc, and the O1 configuration for the noise spectral density which is roughly a factor 2 worse 
than the design sensitivity, we find $\Delta f_{22}\sim2.9\%$, consistent with the error quoted in 
\cite{TheLIGOScientific:2016src}, i.e. $\sigma_{f_{22}}/f_{22}\sim 3.2\%$.}. 

This picture changes for the quality factor, which in general can be measured with smaller precision. The 
right panel of Fig.~\ref{fig:fQ22v2} shows that stellar mass BHs with $M\lesssim 40M_\odot$ always lead to 
errors bigger than $20\%$. In order to determine $Q_{22}$ with a relative accuracy $\Delta Q_{22}<10\%$ 
we need high SNR ringdown events, namely $\rho\gg16$, which lie within the high mass regime. 
Moreover, unlike the mode frequency, the error on the quality factor deteriorates as $j$ increases and 
shows a stronger correlation with the BH final spin, especially for $j\gtrsim 0.6$. 

\begin{figure}[htp]
\centering
\includegraphics[width=4.55cm]{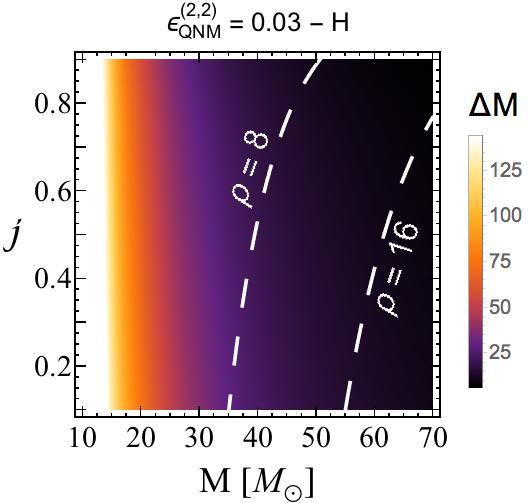}
\includegraphics[width=3.8cm]{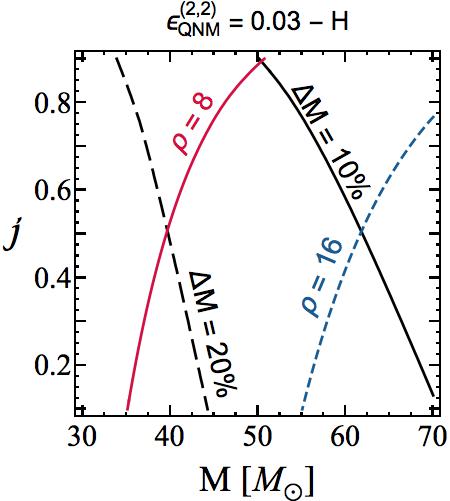}\\
\includegraphics[width=4.55cm]{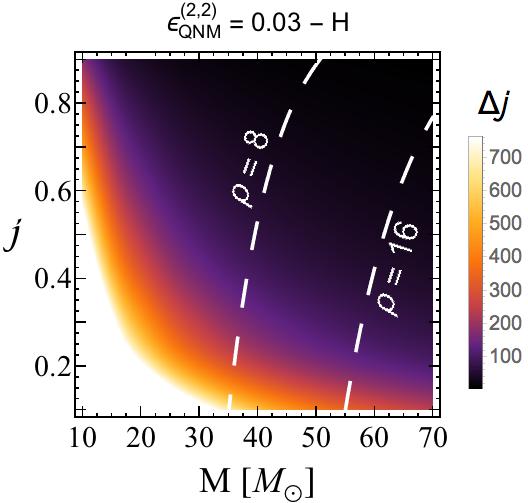}
\includegraphics[width=3.8cm]{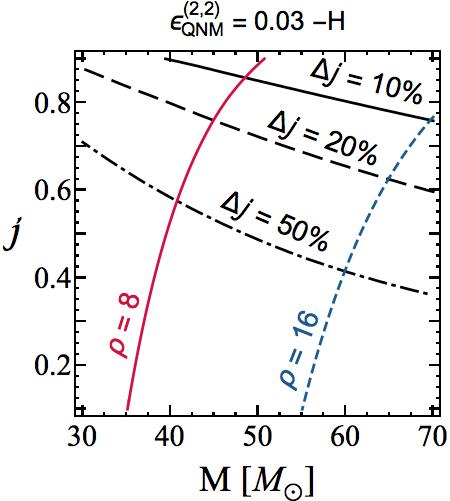}
\caption{Same as Fig.~\ref{fig:fQ22}-\ref{fig:fQ22v2} but for the relative errors on the BH 
mass $M$ and spin parameter $j$, derived from $\Delta{f_{22}}$ and $\Delta Q_{22}$.}
\label{fig:Mj22}
\end{figure}

As discussed in the previous section, the semi-analytic relations \eqref{fitfQ} allow to 
directly compute the error of mass and the angular momentum of the final BH from its ringdown 
frequency and damping time\footnote{As pointed out in \cite{PhysRevD.73.064030} single mode 
detection is not able to uniquely bound the values of mass and spin, as there are multiple 
combinations of $(M,j,l,m,n$) yielding the same frequency and damping time.}, supplied by 
the coordinate transformation \eqref{changebasis}. The values of these uncertainties 
are shown in Fig.~\ref{fig:Mj22}. 

From the left contour plots we immediately note that 
the relative errors become looser. In particular, for the BH mass we find $100\%\lesssim\Delta M\lesssim 10\%$, 
while for the spin  this picture worsens, and we are able to constrain $j$ at the level of $50\%$ or less in a 
limited portion of the parameter space only.
The right panels show again that for a fixed $j$ the error $\Delta M$ decrease as the BH mass grows. 
This is also true if we fix the $M$ and we move to larger values of the spin. However, accuracy of the order of percent 
would require more massive or maximally spinning objects. The data show a strong dependence between $M$ and $j$, 
with correlation coefficients $c_{Mj}>0.98$ for all the configuration considered. As pointed out in 
\cite{Echeverria:1989hg}, this feature would be helpful to determine through gravitational wave observations 
one of the two quantities, as the other has been measured independently by a distinct experiment. 
The right panel of the figure shows that thight constraints on $j$ seem more difficult to achieve, unless 
we consider rapidly rotating BH. Indeed, below $j\sim 0.6$, none of the configurations considered 
yield $\Delta j \lesssim 20\%$, regardless of the body mass.
Finally, as noted at the beginning of this Section, the uncertainties scale linearly with the 
source distance. As a consequence, binary systems at $d=100$ Mpc would translate the 
solid black curves of Fig.~\ref{fig:Mj22} into $\Delta M \sim \Delta j\sim 2\%$.

\begin{figure}[ht]
\centering
\includegraphics[width=4.25cm]{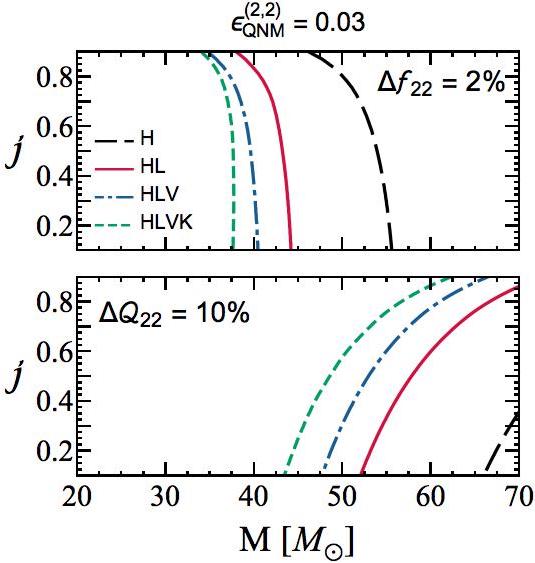}
\includegraphics[width=4.25cm]{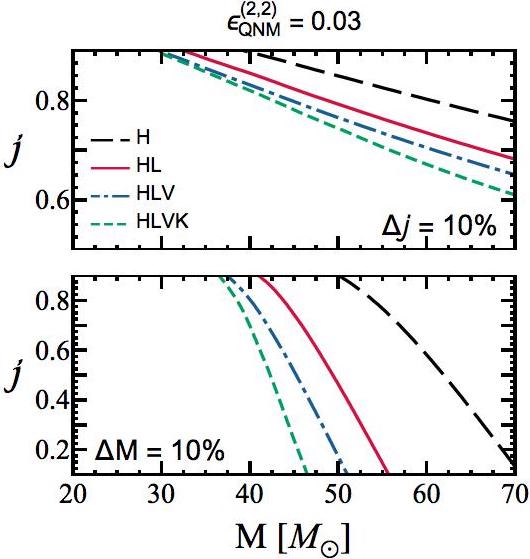}
\caption{Contour plots for fixed relative errors on the frequency/quality factor (left), 
spin/mass (right), as a function of the number of interferometers, for the optimistic 
scenario $\epsilon_\tn{QNM}=0.03$.}
\label{fig:network}
\end{figure}

It is interesting to understand how the previous bounds improve, as we consider multiple 
detectors observing the gravitational wave signal simultaneously. In the left panel of Fig.~\ref{fig:network} we 
show contour lines for fixed errors on the frequency and quality factor, $\Delta f_{22}=2\%$ 
and $\Delta Q_{22}=10\%$, with increasing number of interferometers. As a neat result, in both cases 
the curves move to a lower mass range, making the parameter space available for measurements 
of higher accuracy. This is particularly evident for the quality factor, where configurations with 
$M>40M_\odot$ allow to reach relative errors smaller than $10\%$.  
In the right panel of the same figure we show how these changes reflect into the BH 
mass and spin. Although $\Delta M$ seems to improve more significantly, the bound on 
$j$ remains still quite loose: even with four detectors, accuracy of $10\%$ would require 
spin parameters larger than $0.6$ and massive objects.

All the results derived so far can be immediately translated into the pessimistic scenario 
with $\epsilon_\tn{QNM}=0.01$. Indeed, the radiation efficiency acts as a multiplicative 
factor for the gravitational wave amplitude, which makes the relative errors proportional to\footnote{Viceversa the SNR scales as $\rho \propto \epsilon_\tn{QNM}$.} $1/\epsilon_\tn{QNM}$.
As a golden rule, the uncertainties can be simply rescaled depending on the overall amount of 
energy stored into the modes. 
In particular, in our case the bounds weaken, with the errors increasing  by a factor of $3$. 
This is particularly penalizing for the quality factor, as we wouldn't have $\Delta Q_{22}\lesssim 50\%$ 
for BH lighter than $45 M_\odot$.

\subsection{Second generation detectors: multiple modes}

Future detections of a secondary mode would represent a precious source of information on the nature of 
the compact object, especially as far as strong-field tests of the no-hair theorem 
are concerned. Therefore, in this section we focus on the detectability of the $(3,3)$ and $(4,4)$ 
QNMs, which provide the largest contribution 
to the total GW energy budget, regardless of the binary progenitor's spins. 
Relativistic numerical simulations show indeed that the mode's hierarchy is mostly 
affected by the mass ratio of the inspiral phase and by the source orientation 
\cite{PhysRevD.75.124018,PhysRevD.76.064034,Healy:2013jza}. 
Moreover, we consider physical scenarios in which the second mode carries a fraction $10^{-1}$ 
of the energy radiated by the $(2,2)$ component. 

\begin{figure*}[ht]
\centering
\hspace{0.15cm}
\includegraphics[width=5.1cm]{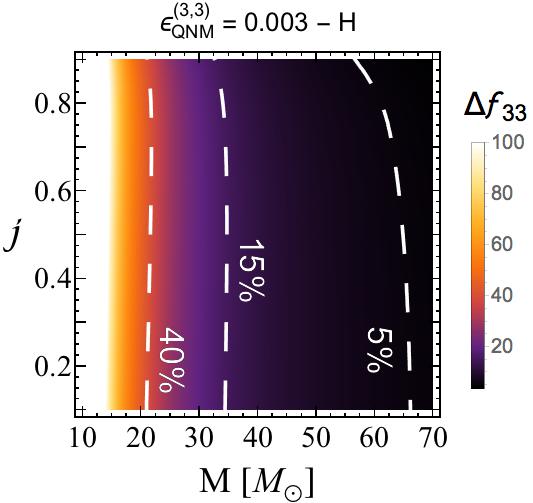}
\includegraphics[width=5.1cm]{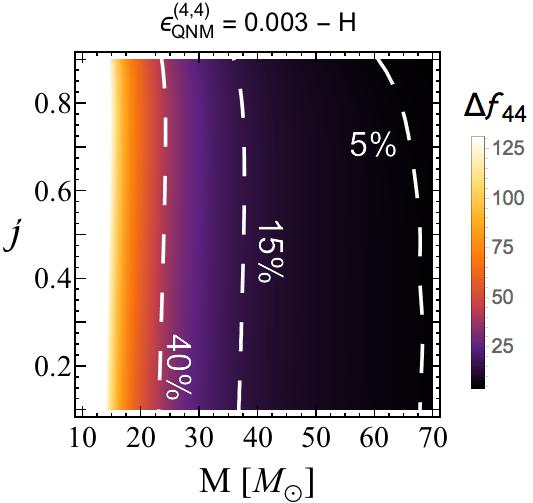}\\
\vspace{0.2cm}
\includegraphics[width=4.3cm]{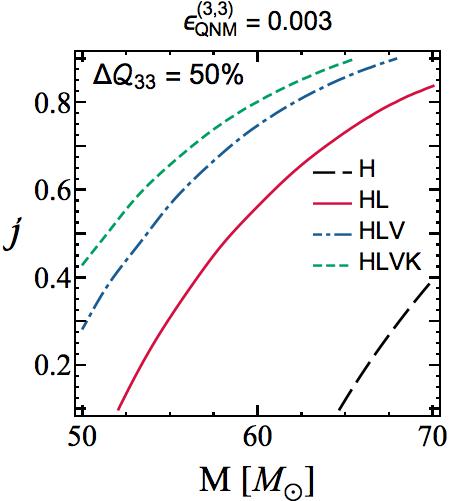}
\includegraphics[width=4.3cm]{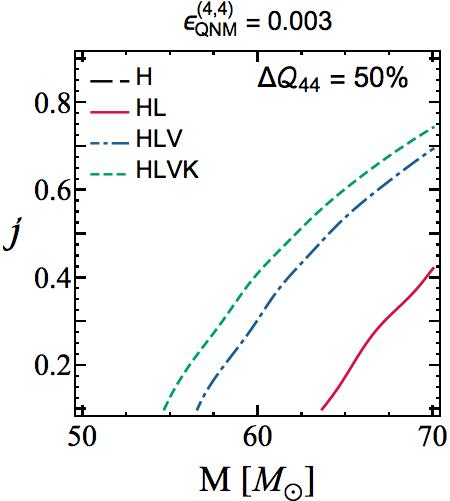}
\includegraphics[width=4.3cm]{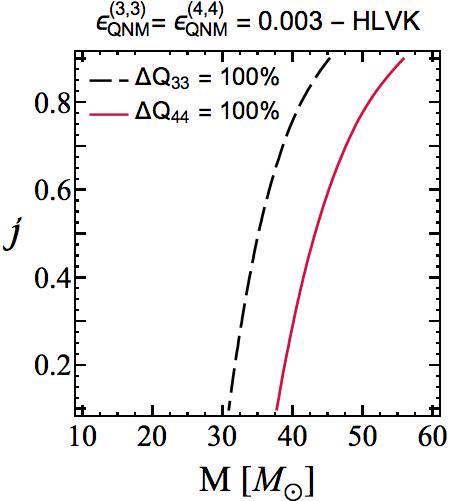}
\hspace{0.2cm}
\caption{(Top) Contour plots of relative uncertainty for the frequency of 
the $(3,3)$ and $(4,4)$ modes, with $\epsilon^{(3,3)}_\tn{QNM}=\epsilon^{(4,4)}_\tn{QNM}=0.003$. 
(Bottom) Curves of fixed accuracy $\Delta Q_{33}=\Delta Q_{44}=(50,100)\%$ for a network of interferometers 
with increasing number of detectors.}
\label{fig:fQ3344}
\end{figure*}

In the top row of Fig.~\ref{fig:fQ3344} we show the BH configurations leading to fixed errors 
on $f_{33}$ and $f_{44}$ as a function of mass and spin, for our optimistic scenario. The two 
figures yield some similar features, regardless of the specific choice of $(l,m)$. 
As expected, the relative errors worsen, as a result of the lower total energy stored into 
the modes, and the higher frequencies\footnote{For a fixed $j$ the mode frequency $f_{lm}$ increases with $l$.}. In this case, 
for the same BH models shown in Fig.~\ref{fig:fQ22}, $\Delta f_{33}$ and $\Delta f_{44}$ are around $\sim3\div5$ and 
$\sim5\div 7$ times larger than the uncertainties of the $(2,2)$ mode, respectively. This is more evident for the term with 
$l=m=4$, for which higher accuracy measurements shift to the end of the mass and spin ranges, $M\sim 70M_\odot$ 
and $j\gtrsim 0.7$.
Moreover, within the parameter space considered, the SNR of the secondary modes are always 
$[80\div90]\%$ smaller of the SNR obtained for the $(2,2)$ component. 
These changes are more dramatic for the quality factor, as the bottom row of Fig.~\ref{fig:fQ3344} 
shows that it would even be difficult for AdLIGO {\it alone} to set a bound $\Delta Q_{33}=50\%$. 
Accuracy of the order of $\Delta Q_{33}\lesssim 40\%$ would require 
BH configurations out of the considered parameter space. As seen for the single mode analysis, these bounds 
improve when a network of interferometers is considered. With four instruments at design 
sensitivity we would roughly gain a factor of 2, shifting the constraint $\Delta Q_{33}=50\%$
to lower masses, and allowing the parameter space to set narrower constraints.
Multiple detections would be particularly relevant for the $(4,4)$ case, since with only one LIGO 
site, all the prototype BHs analyzed lead to uncertainties $\Delta Q_{44}\gg50\%$. However, from the last bottom 
panel of Fig.~\ref{fig:fQ3344} we note that for both the $(3,3)$ and the $(4,4)$ modes even four detectors would provide poor 
bounds, leaving the quality factors unconstrained, i.e. $\Delta Q_{33}=\Delta Q_{44}>100\%$, in a significant portion of the parameter space. 

As a final comment, we note that the relative uncertainties presented in 
Fig.~\ref{fig:fQ3344} would strongly deteriorate for the astrophysical scenario in which 
$\epsilon^{(3,3)}_\tn{QNM}=\epsilon^{(2,2)}_\tn{QNM}/10=0.001$, increasing, as 
discussed before, by a factor of 3. Although still loosing sensitive to the QNM frequency, 
the pessimistic assumption would make the interferometers essentially blind to the quality 
factor of the second mode.

It is worth to remark that before discussing the errors on the secondary modes, we should 
have clarified whether the two components can be distinguished within the noise. As already discussed
in literature, we follow a two-criterion analysis \cite{Bhagwat:2016ntk,PhysRevD.73.064030}.

\begin{figure}[ht]
\centering
\includegraphics[width=4.25cm]{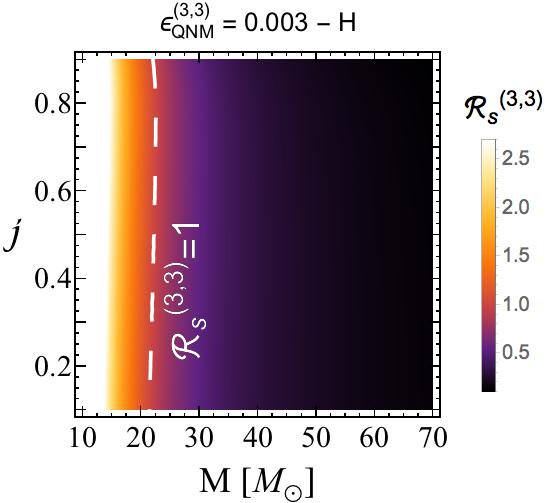}
\includegraphics[width=4.25cm]{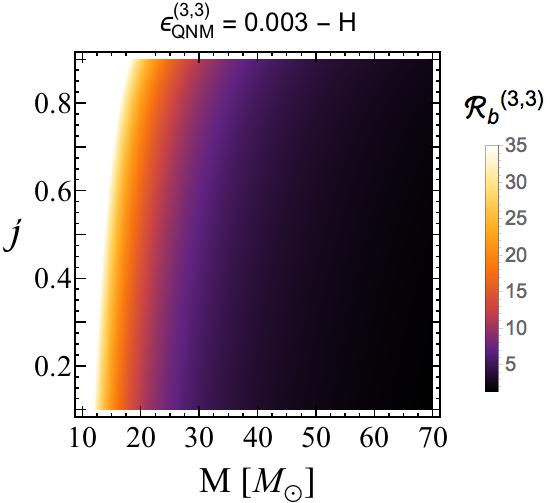}
\caption{Contour plots for the critical values ${\cal R}_{s}$ and ${\cal R}_b$ required 
to distinguish frequencies and quality factors of the $(2,2)$ and the $(3,3)$ modes.}
\label{fig:rhoeff}
\end{figure}

First, we identify a threshold SNR, which each QNM has to exceed in order to be visible. 
We set this value to $\rho_\tn{th}=5$. Then, in order to be able to disentangle the features between 
two modes characterized by frequencies and quality factors $(f_1,f_2)$ and $(Q_1,Q_2)$ we 
employ a Rayleigh criterion, which specifically introduce the critical SNR ratios
\begin{align}
\frac{\rho^f_\tn{cr}}{\rho^{(l,m)}}=&\frac{\max(\sigma_{f_1}, \sigma_{f_2})}{\vert f_1-f_2\vert}\ ,\\
\frac{\rho^Q_\tn{cr}}{\rho^{(l,m)}}=&\frac{\max( \sigma_{Q_1}, \sigma_{Q_2})}{\vert Q_1-Q_2\vert}\ ,
\end{align}
where $\rho^{(l,m)}$ refers to the single component. In our analysis $f_1=f_{22},Q_1=Q_{22}$, while 
$(f_2,Q_2)$ correspond either to the $(3,3)$ or the $(4,4)$ term.
To distinguish the frequency or the quality factor between the modes we ask that
\begin{equation}\label{rhoc1}
{\cal R}^{(l,m)}_s=\min\left(\frac{\rho^f_\tn{cr}}{\rho^{(l,m)}},\frac{\rho^Q_\tn{cr}}{\rho^{(l,m)}}\right)<1\ ,
\end{equation}
while, to resolve both 
\begin{equation}\label{rhoc2}
{\cal R}^{(l,m)}_b=\max\left(\frac{\rho^f_\tn{cr}}{\rho^{(l,m)}},\frac{\rho^Q_\tn{cr}}{\rho^{(l,m)}}\right)<1\ .
\end{equation}
The ratios between the SNR for the $(3,3)$ mode and the critical values ${\cal R}^{(3,3)}_s$ and ${\cal R}^{(3,3)}_b$ 
are plotted in Fig.~\ref{fig:rhoeff}. The right panel shows that the second condition ${\cal R}_b<1$ is never 
satisfied by the considered configurations, and therefore AdLIGO alone would not be able to resolve both 
$f_{lm}$ and $Q_{lm}$. Nevertheless, looking at the left panel, it seems that for $M> 20M_\odot$ the two QNM 
frequencies (or quality factors) can always be distinguished, as ${\cal R}_s<1$. 

\begin{figure}[ht]
\centering
\includegraphics[width=4.25cm]{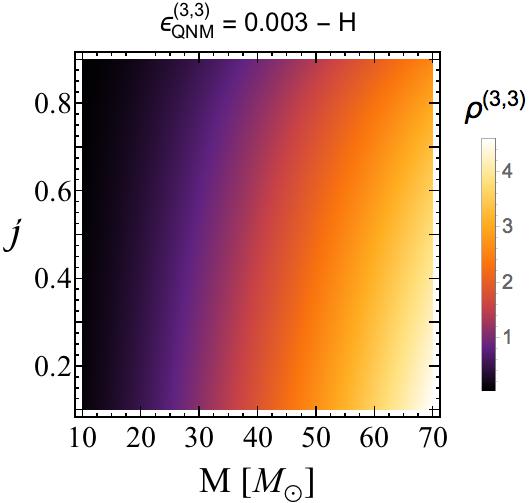}
\includegraphics[width=4.25cm]{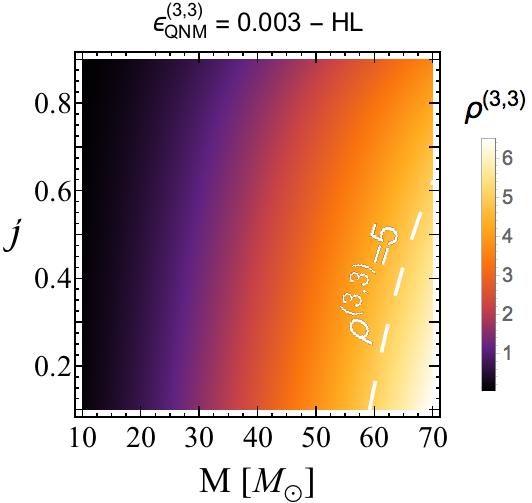}\\
\includegraphics[width=4.25cm]{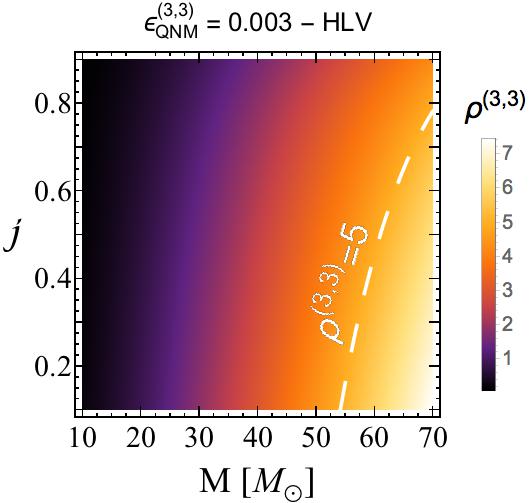}
\includegraphics[width=4.25cm]{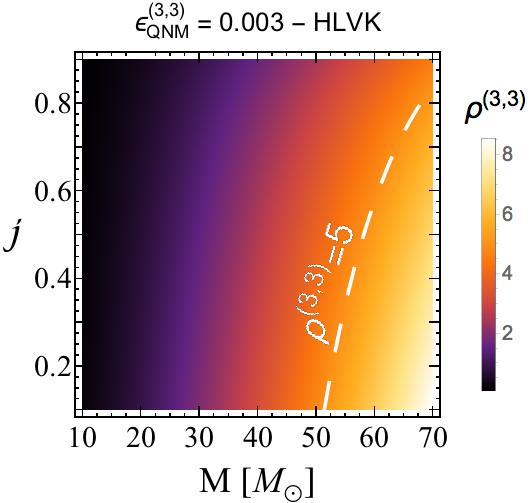}
\caption{Change in the SNR for the secondary mode $\rho^{(3,3)}$, as a function of the number of  
terrestrial interferometers. The white dashed curve correspond to the threshold 
$\rho^{(3,3)}=5$.}
\label{fig:rhoeff2}
\end{figure}

However, as shown in the top-left panel of Fig.~\ref{fig:rhoeff2}, a computation of the $(3,3)$ SNR points 
out that none of the BH configurations satisfies the first detectability condition 
$\rho^{(3,3)}\geq \rho_\tn{th}=5$. This results confirm the statistical analysis carried out 
in \cite{Berti:2016lat}, suggesting that multimode analysis seems unfeasible with only one 
interferometer.
Overall, these data do not change qualitatively if we consider as secondary mode the $(4,4)$ 
component.

The prospect of BH spectroscopy increases as far as a network of detector will be operational. 
The remaining three contour plots of Fig.~\ref{fig:rhoeff2} show the evolution of $\rho^{(3,3)}$ as a 
function of the number of interferometers. We immediately note how the white-dashed curve, which identifies 
the BH configuration leading to $\rho_\tn{th}=5$, shifts within the parameter space. Such enhancement allows 
to analyze ringdown events  which fulfil the threshold condition $\rho^{(3,3)}\geq5$. 

\begin{figure}[ht]
\centering
\includegraphics[width=4.25cm]{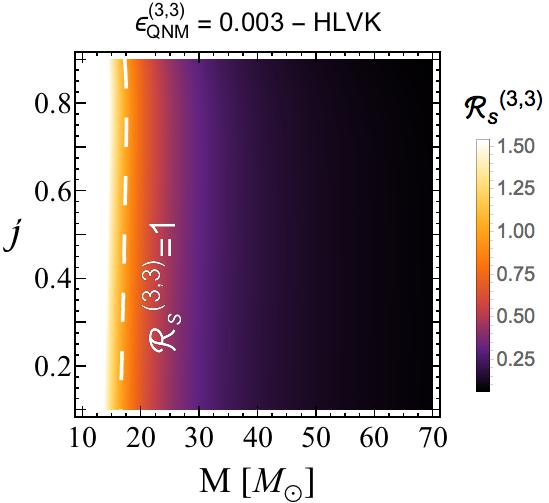}
\includegraphics[width=4.25cm]{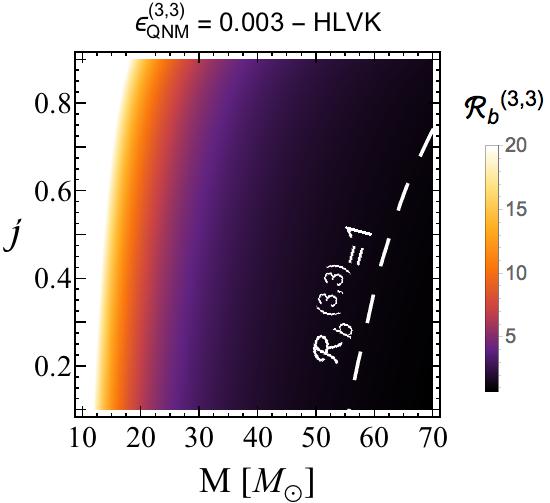}
\caption{Same as Fig.~\ref{fig:rhoeff} but for the case of four interferometers observing the 
ringdown event.}
\label{fig:rhoeff3}
\end{figure}

At the same time, the critical ratios \eqref{rhoc1}-\eqref{rhoc2} also improve, as shown in  
Fig.~\ref{fig:rhoeff3}, where for sake of clarity we only consider the case with four detectors at 
design sensitivity. Beside a small change of the condition ${\cal R}_{s}=1$ to BH masses smaller 
than $20M_\odot$, the most interesting result is that for $M\gtrsim 55M_\odot$ also the second 
criterion, ${\cal R}_b^{(3,3)}<1$, is satisfied. Therefore, a complete synergy between the current 
and near-future facilities is crucial to fully exploit BH spectroscopy to probe the validity of the 
no-hair theorem.

As final remark, we note that in the less optimistic scenario with energy radiated 
$\epsilon^{(3,3)}_\tn{QNM}=\epsilon^{(2,2)}_\tn{QNM}/10=0.01$, even with four interferometers, 
all the BHs analyzed would provide $\rho^{(3,3)}<5$, making the detection of the subdominant 
mode, and therefore strong-field tests of GR, extremely more difficult to achieve. However, more 
sophisticate data analysis techniques would also improve this picture. Indeed, as recently shown in 
\cite{Yang:2017zxs}, coherent stacking of several ringdown events would significantly increase the 
SNR of the secondary mode, leading therefore to a precise estimate of its fundamental parameters.
Finally, it is worth to mention that in some cases, even a single mode analysis can constrain the 
nature of the ringing compact object, as described in \cite{Chirenti:2016hzd}, in which the QNMs 
properties obtained from GW150914 have been used to exclude the existence of a rotating gravastar.

\subsection{Third generation detectors} \label{Sec:future}

Being the main focus our work to explore the constraints that current detectors, like LIGO 
and Virgo, may be able to set on the BH ringdown, it is interesting to analyze how 
these results improve as far as  third generation of interferometers are considered. In this 
section we shall focus on four new instruments: (i) LIGO A+ and LIGO Voyager, 
which represent  a major upgrade of the existing facilities with frequency-dependent squeezed 
light, improved mirrors and laser beams, (ii) the Cosmic Explorer, which 
is essentially a new detector with $40$ km arm-length based on the LIGO A+ technology, (iii) the 
European Einstein Telescope, with three arms of $10$ km in a triangle configuration  built 
underground \cite{Clark:2015zxa}.

\begin{figure*}[ht]
\centering
\includegraphics[width=4.5cm]{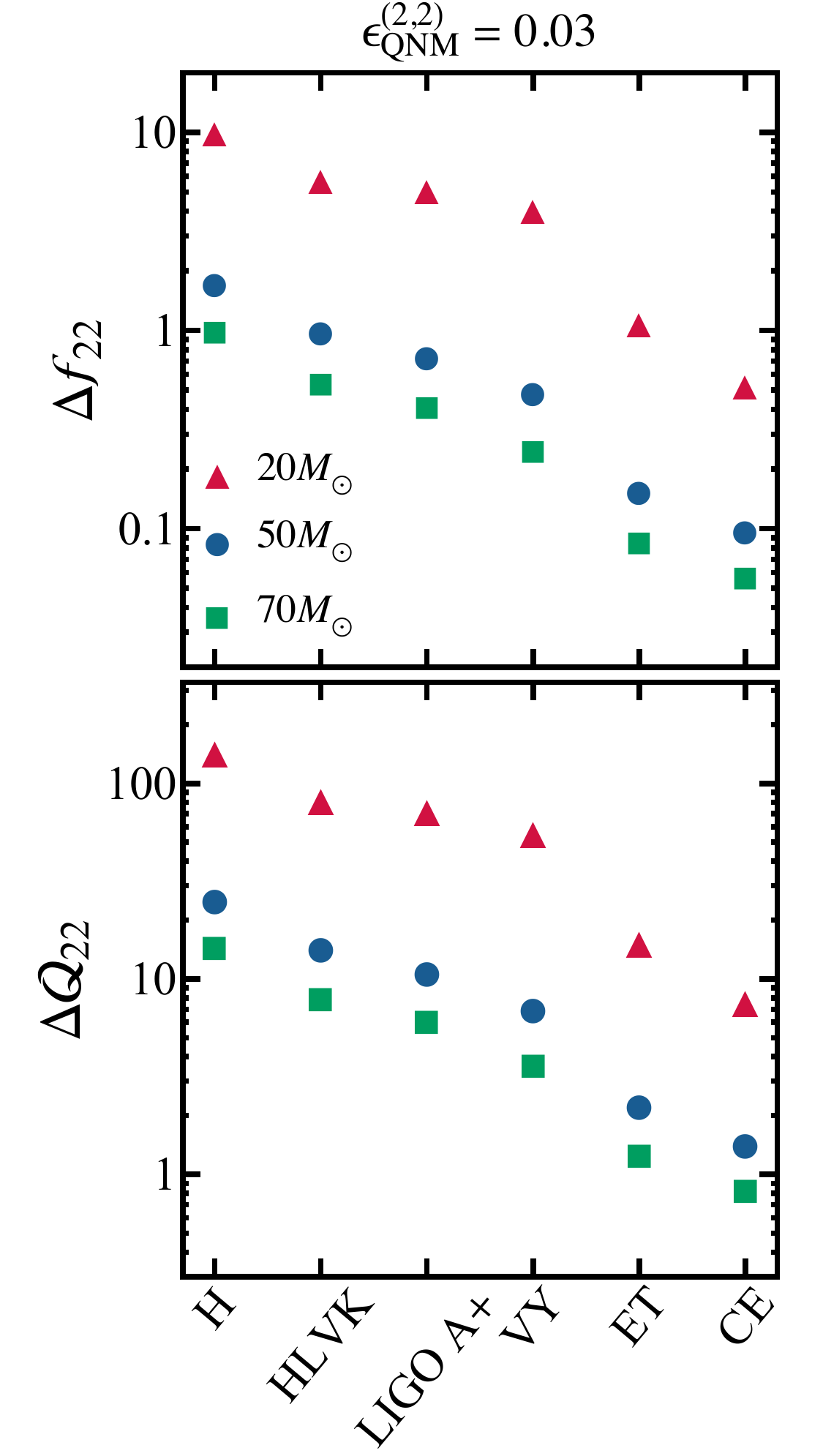}
\includegraphics[width=4.5cm]{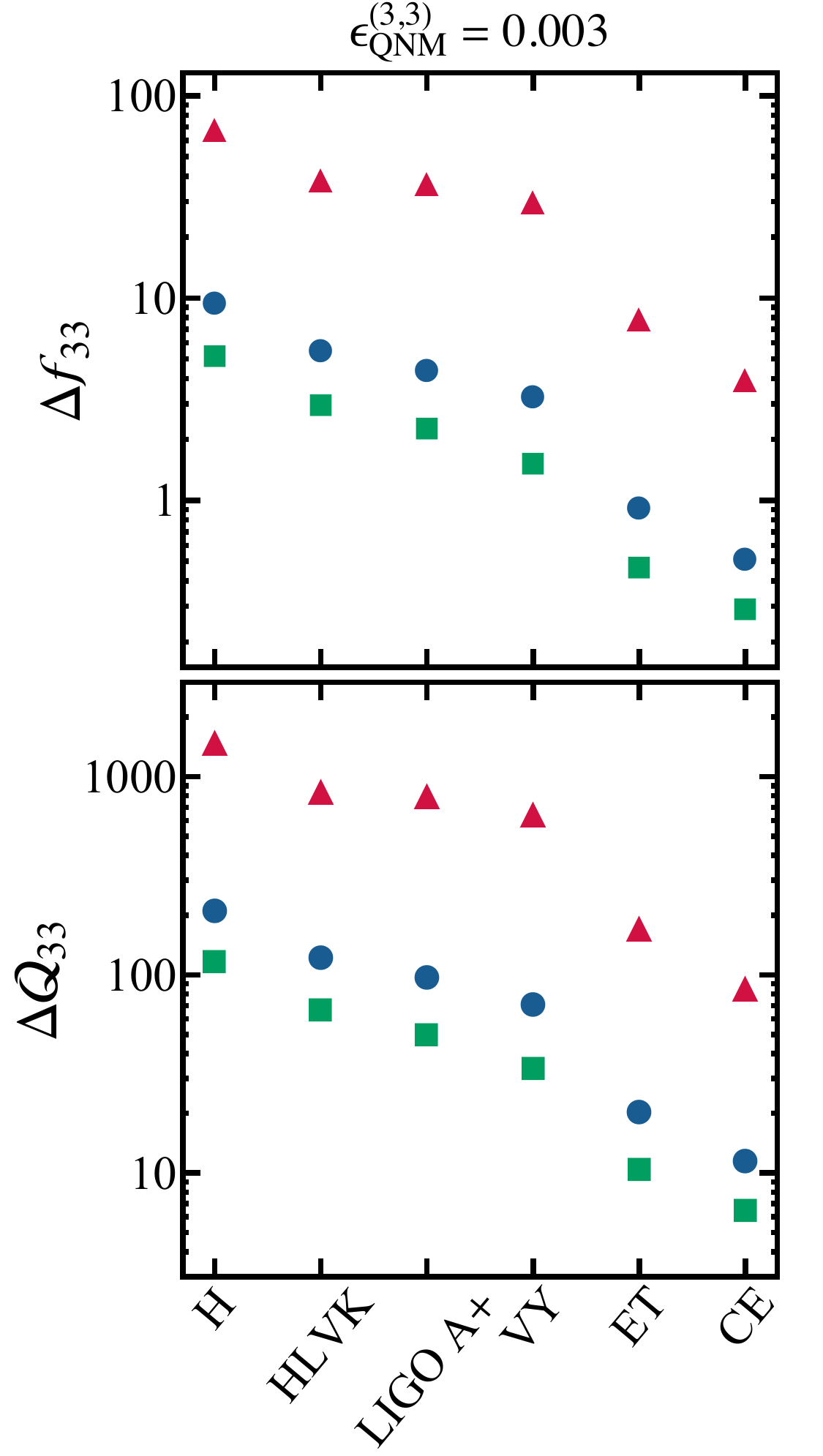}
\includegraphics[width=4.5cm]{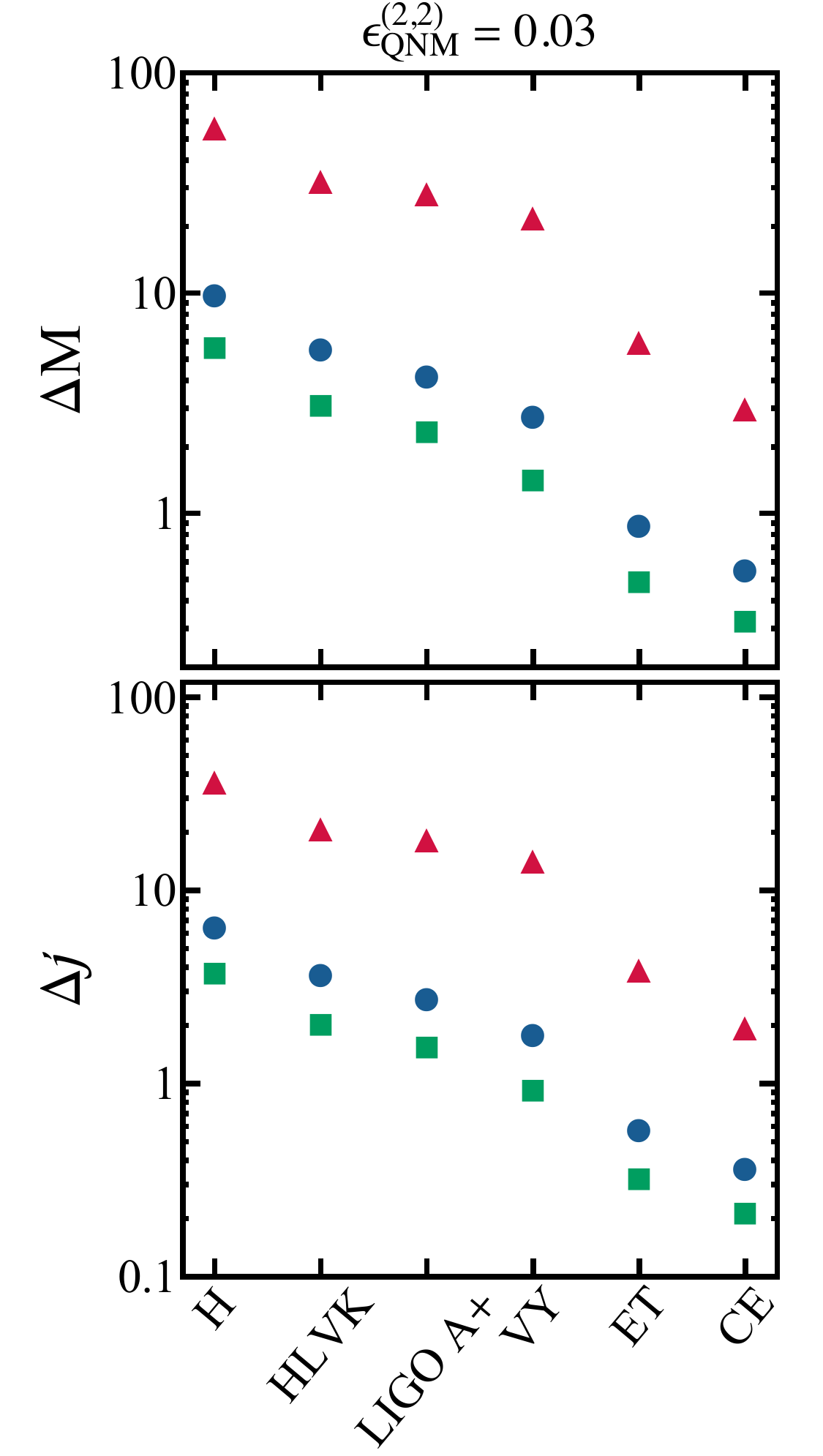}
\caption{Left and center panels show the relative errors on frequency and quality factor of the $(2,2)$ 
and the $(3,3)$ modes, for second and third generation of interferometers. Uncertainties on the mass and 
spin computed from the dominant $(2,2)$ component are drawn in the right plot. We consider three mass 
configurations $M=(20,50,70)M_\odot$, assuming BHs with spin parameter $j=0.9$.}
\label{fig:future}
\end{figure*}

A direct comparison on the accuracies of ringdown parameters measured with current and 
future interferometers is shown in Fig.~\ref{fig:future}, for both single and multi-mode events. 
For the latter, we only analyze the $(3,3)$ component, although our results will also 
apply for the $(4,4)$ term. We consider three classes of {\it light}, {\it medium} and {\it heavy} 
BHs, with masses of $20M_\odot$, $50M_\odot$ and $70M_\odot$, respectively. For sake of 
simplicity, we also assume a best-case scenario with spin parameter $j=0.9$, and radiated 
energy $\epsilon^{(2,2)}_\tn{QNM}=0.3 $. 
Left and center panels of Fig.~\ref{fig:future} yield several new considerations on the relative 
errors of the mode's frequencies and damping times. Interestingly, for all the configurations 
a network of 4 second generation detectors perform almost as well as LIGO A+. A Voyager class 
mission will not lead to major improvements, as well. However,  the Einstein 
telescope and the Cosmic Explorer changes this picture dramatically. This is particularly evident for 
massive sources, with $M\gtrsim 50M_\odot$, for which we expect to measure both $f_{22}$ and 
$Q_{22}$ with relative accuracy better than $1\%$. Such results are even more relevant for the quality 
factor of the secondary component. In this case LIGO A+ will be able to set an upper bound, 
$\Delta Q_{33}=100\%$, for BHs with $M\sim50M_\odot$, but only ET and CE are potentially able to 
constrain lighter sources. The exquisite sensitivity of the Cosmic Explorer is indeed necessary 
to provide useful measurements for BHs with $M\lesssim20M_\odot$. 
These results directly translate on the errors on mass and spin parameter, as shown in the right 
panel of Fig.~\ref{fig:future}. For both these quantities ET and CE are the only future detectors 
with projected constraints smaller than $10\%$.

\begin{figure}[ht]
\centering
\includegraphics[width=6cm]{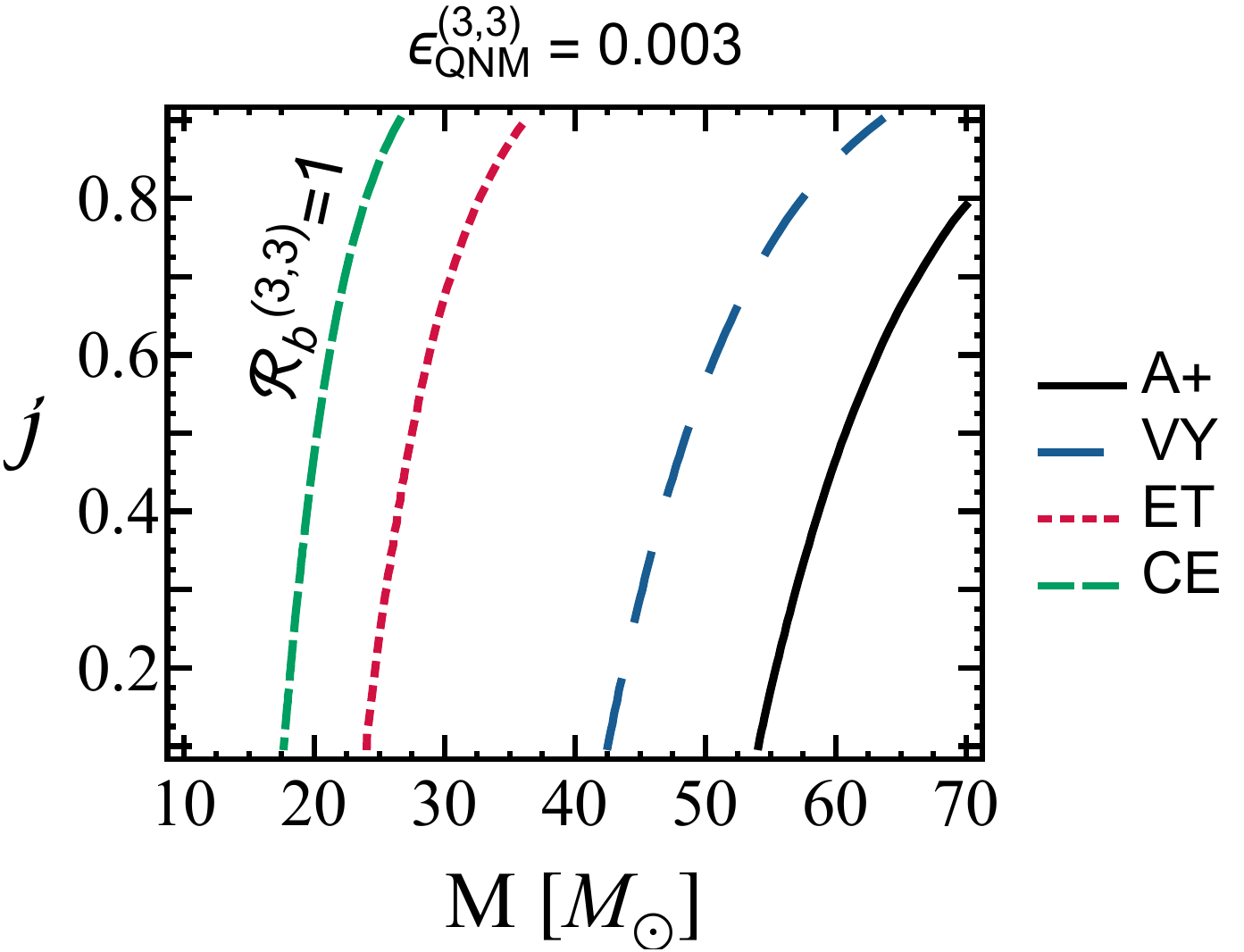}
\caption{Contour plots for the critical value ${\cal R}_b$ need 
to distinguish both the frequency and the quality factor of the $(2,2)$ and the $(3,3)$ 
modes.}
\label{fig:future2}
\end{figure}

As a final comment, in Fig.~\ref{fig:future2} we show contour lines of fixed ${\cal R}_\tn{b}$,
which correspond to the critical value needed to distinguish both the frequency and quality factor of the 
$(2,2)$ and the $(3,3)$ components. Comparing this results with the right panel of Fig.~\ref{fig:rhoeff3}, 
we immediately note that even in this case the network configuration behaves as LIGO A+. 
As expected, best results are obtained for ET and CE, which are able to resolve both modes 
within almost all the parameter space.

\section{Conclusions} \label{Sec:conc}

With the LIGO observatories already operational and Virgo in the final completion phase, 
gravitational wave astronomy represents an exciting and active field of research. 
Advanced interferometers, supplied by the expected detection rates \cite{Dominik:2012kk,Dominik:2014yma}, 
promise to detect compact binary coalescences as a weekly routine, leading to a new flood of data to 
be analyzed. 
Astrophysical BHs produced by such systems are ideal candidates to investigate gravity in extreme 
conditions, and to test the validity of GR foundations, as the no-hair theorem, through emission 
of QNMs. 

In this paper we have analyzed the detectability of such signals. Using a Fisher matrix approach, we have 
spanned the BH-parameter space, focusing on the constraints that current and near future terrestrial detectors 
will be able to set on the ringdown parameters. For second generation of interferometers, 
our numerical results can be summarized as follows.
\begin{itemize}
\item For a single mode detection, in a favorable scenario with radiated energy $\epsilon_\tn{QNM}=0.03$, 
frequency and quality factor of the fundamental $(2,2)$ mode can be measured with relative accuracy of 
$20\%\lesssim\Delta f_{22}\lesssim1\%$ and $100\%\lesssim\Delta Q_{22}\lesssim20\%$, respectively. 
A pessimistic assumption with $\epsilon_\tn{QNM}=0.01$ would increase these errors by a factor of 3.
\item These values impact on our ability to constrain mass and spin of the newly born BH. Relative errors 
of $20\%\lesssim \Delta M\lesssim 10\%$ are feasible for $M\gtrsim 40M_\odot$ only, while for the angular 
momentum we obtain $50\%\lesssim \Delta j\lesssim 10\%$ for $j>0.4$ only, almost independently from the BH 
final mass.
\item The previous estimates improve with multiple detectors. Four interferometers at design sensitivity would 
reduce the relative errors roughly by a factor of 2.
\item AdLIGO alone is not able to perform test of GR with a single ringdown observation. For the configurations 
considered in this paper, all the events yield secondary modes with $l=m=3$ and $l=m=4$ 
under the SNR threshold $\rho_\tn{th}=5$. This would make impossible to extract the subdominant 
components from the instrument's noise.
\item A network of detectors at full sensitivity will be able to fully exploit the multi mode analysis distinguishing 
both the frequency and the quality factor of the first and second QNMs, allowing for strong field tests of the no-hair 
theorem.
\end{itemize}

These results improve for third generation detectors, which we have included in our analysis for a direct comparison 
with the current facilities. For all the BH considered, the LIGO A+ upgrade has performances similar to the network 
HLVK. Major changes occur for the Einstein Telescope and the Cosmic Explorer, which represent the best instruments 
to constrain both frequencies and damping times of the dominant (2,2) mode, with 
accuracy at the level of $1\%$ and better. Such results also extend to the secondary component, as CE is the only 
interferometer potentially able to set an upper bound on the $(3,3)$ mode for less massive sources with 
$M\lesssim 20M_\odot$.

\begin{acknowledgments}
A.M. acknowledges Claudia Lazzaro and Irene di Palma for useful and stimulating discussions. The authors 
are grateful to Emanuele Berti and Matthew Evans for sharing the numerical data of the sensitivity curves 
of third generation interferometers. P. L. is supported by the NSF Grants No. 1505824 and No. 1333360.
\end{acknowledgments}

\bibliography{bibnote}

\begin{thebibliography}{50}
\expandafter\ifx\csname natexlab\endcsname\relax\def\natexlab#1{#1}\fi
\expandafter\ifx\csname bibnamefont\endcsname\relax
  \def\bibnamefont#1{#1}\fi
\expandafter\ifx\csname bibfnamefont\endcsname\relax
  \def\bibfnamefont#1{#1}\fi
\expandafter\ifx\csname citenamefont\endcsname\relax
  \def\citenamefont#1{#1}\fi
\expandafter\ifx\csname url\endcsname\relax
  \def\url#1{\texttt{#1}}\fi
\expandafter\ifx\csname urlprefix\endcsname\relax\def\urlprefix{URL }\fi
\providecommand{\bibinfo}[2]{#2}
\providecommand{\eprint}[2][]{\url{#2}}

\bibitem[{\citenamefont{Abbott et~al.}(2016{\natexlab{a}})}]{Abbott:2016blz}
\bibinfo{author}{\bibfnamefont{B.~P.} \bibnamefont{Abbott}}
  \bibnamefont{et~al.} (\bibinfo{collaboration}{Virgo, LIGO Scientific}),
  \bibinfo{journal}{Phys. Rev. Lett.} \textbf{\bibinfo{volume}{116}},
  \bibinfo{pages}{061102} (\bibinfo{year}{2016}{\natexlab{a}}),
  \eprint{1602.03837}.

\bibitem[{\citenamefont{Abbott et~al.}(2016{\natexlab{b}})}]{Abbott:2016nmj}
\bibinfo{author}{\bibfnamefont{B.~P.} \bibnamefont{Abbott}}
  \bibnamefont{et~al.} (\bibinfo{collaboration}{Virgo, LIGO Scientific}),
  \bibinfo{journal}{Phys. Rev. Lett.} \textbf{\bibinfo{volume}{116}},
  \bibinfo{pages}{241103} (\bibinfo{year}{2016}{\natexlab{b}}),
  \eprint{1606.04855}.

\bibitem[{\citenamefont{Kokkotas and Schmidt}(1999)}]{Kokkotas:1999bd}
\bibinfo{author}{\bibfnamefont{K.~D.} \bibnamefont{Kokkotas}} \bibnamefont{and}
  \bibinfo{author}{\bibfnamefont{B.~G.} \bibnamefont{Schmidt}},
  \bibinfo{journal}{Living Rev. Rel.} \textbf{\bibinfo{volume}{2}},
  \bibinfo{pages}{2} (\bibinfo{year}{1999}), \eprint{gr-qc/9909058}.

\bibitem[{\citenamefont{Vishveshwara}(1970)}]{Vishveshwara:1970zz}
\bibinfo{author}{\bibfnamefont{C.~V.} \bibnamefont{Vishveshwara}},
  \bibinfo{journal}{Nature} \textbf{\bibinfo{volume}{227}},
  \bibinfo{pages}{936} (\bibinfo{year}{1970}).

\bibitem[{\citenamefont{Teukolsky}(1973)}]{Teukolsky:1973ha}
\bibinfo{author}{\bibfnamefont{S.~A.} \bibnamefont{Teukolsky}},
  \bibinfo{journal}{Astrophys. J.} \textbf{\bibinfo{volume}{185}},
  \bibinfo{pages}{635} (\bibinfo{year}{1973}).

\bibitem[{\citenamefont{Chandrasekhar and
  Detweiler}(1975)}]{Chandrasekhar:1975zza}
\bibinfo{author}{\bibfnamefont{S.}~\bibnamefont{Chandrasekhar}}
  \bibnamefont{and} \bibinfo{author}{\bibfnamefont{S.~L.}
  \bibnamefont{Detweiler}}, \bibinfo{journal}{Proc. Roy. Soc. Lond.}
  \textbf{\bibinfo{volume}{A344}}, \bibinfo{pages}{441} (\bibinfo{year}{1975}).

\bibitem[{\citenamefont{Price and Pullin}(1994)}]{Price:1994pm}
\bibinfo{author}{\bibfnamefont{R.~H.} \bibnamefont{Price}} \bibnamefont{and}
  \bibinfo{author}{\bibfnamefont{J.}~\bibnamefont{Pullin}},
  \bibinfo{journal}{Phys. Rev. Lett.} \textbf{\bibinfo{volume}{72}},
  \bibinfo{pages}{3297} (\bibinfo{year}{1994}), \eprint{gr-qc/9402039}.

\bibitem[{\citenamefont{Wiltshire et~al.}(2009)\citenamefont{Wiltshire, Visser,
  and Scott}}]{wiltshire2009kerr}
\bibinfo{author}{\bibfnamefont{D.~L.} \bibnamefont{Wiltshire}},
  \bibinfo{author}{\bibfnamefont{M.}~\bibnamefont{Visser}}, \bibnamefont{and}
  \bibinfo{author}{\bibfnamefont{S.~M.} \bibnamefont{Scott}},
  \emph{\bibinfo{title}{{The Kerr spacetime: Rotating black holes in general
  relativity}}} (\bibinfo{publisher}{Cambridge University Press},
  \bibinfo{year}{2009}), ISBN \bibinfo{isbn}{9780521885126},
  \urlprefix\url{http://www.cambridge.org/catalogue/catalogue.asp?isbn=9780521885126}.

\bibitem[{\citenamefont{Cardoso and Gualtieri}(2016)}]{Cardoso:2016ryw}
\bibinfo{author}{\bibfnamefont{V.}~\bibnamefont{Cardoso}} \bibnamefont{and}
  \bibinfo{author}{\bibfnamefont{L.}~\bibnamefont{Gualtieri}},
  \bibinfo{journal}{Class. Quant. Grav.} \textbf{\bibinfo{volume}{33}},
  \bibinfo{pages}{174001} (\bibinfo{year}{2016}), \eprint{1607.03133}.

\bibitem[{\citenamefont{Hawking}(1971)}]{PhysRevLett.26.1344}
\bibinfo{author}{\bibfnamefont{S.~W.} \bibnamefont{Hawking}},
  \bibinfo{journal}{Phys. Rev. Lett.} \textbf{\bibinfo{volume}{26}},
  \bibinfo{pages}{1344} (\bibinfo{year}{1971}),
  \urlprefix\url{http://link.aps.org/doi/10.1103/PhysRevLett.26.1344}.

\bibitem[{\citenamefont{Amaro-Seoane et~al.}(2013)}]{AmaroSeoane:2012km}
\bibinfo{author}{\bibfnamefont{P.}~\bibnamefont{Amaro-Seoane}}
  \bibnamefont{et~al.}, \bibinfo{journal}{GW Notes}
  \textbf{\bibinfo{volume}{6}}, \bibinfo{pages}{4} (\bibinfo{year}{2013}),
  \eprint{1201.3621}.

\bibitem[{\citenamefont{Amaro-Seoane et~al.}(2012)}]{AmaroSeoane:2012je}
\bibinfo{author}{\bibfnamefont{P.}~\bibnamefont{Amaro-Seoane}}
  \bibnamefont{et~al.}, \bibinfo{journal}{Class. Quant. Grav.}
  \textbf{\bibinfo{volume}{29}}, \bibinfo{pages}{124016}
  (\bibinfo{year}{2012}), \eprint{1202.0839}.

\bibitem[{\citenamefont{Echeverria}(1989)}]{Echeverria:1989hg}
\bibinfo{author}{\bibfnamefont{F.}~\bibnamefont{Echeverria}},
  \bibinfo{journal}{Phys. Rev.} \textbf{\bibinfo{volume}{D40}},
  \bibinfo{pages}{3194} (\bibinfo{year}{1989}).

\bibitem[{\citenamefont{Flanagan and
  Hughes}(1998{\natexlab{a}})}]{PhysRevD.57.4566}
\bibinfo{author}{\bibfnamefont{E.~E.} \bibnamefont{Flanagan}} \bibnamefont{and}
  \bibinfo{author}{\bibfnamefont{S.~A.} \bibnamefont{Hughes}},
  \bibinfo{journal}{Phys. Rev. D} \textbf{\bibinfo{volume}{57}},
  \bibinfo{pages}{4566} (\bibinfo{year}{1998}{\natexlab{a}}),
  \urlprefix\url{http://link.aps.org/doi/10.1103/PhysRevD.57.4566}.

\bibitem[{\citenamefont{Flanagan and
  Hughes}(1998{\natexlab{b}})}]{Flanagan:1997sx}
\bibinfo{author}{\bibfnamefont{E.~E.} \bibnamefont{Flanagan}} \bibnamefont{and}
  \bibinfo{author}{\bibfnamefont{S.~A.} \bibnamefont{Hughes}},
  \bibinfo{journal}{Phys. Rev.} \textbf{\bibinfo{volume}{D57}},
  \bibinfo{pages}{4535} (\bibinfo{year}{1998}{\natexlab{b}}),
  \eprint{gr-qc/9701039}.

\bibitem[{\citenamefont{Berti et~al.}(2006{\natexlab{a}})\citenamefont{Berti,
  Cardoso, and Will}}]{PhysRevD.73.064030}
\bibinfo{author}{\bibfnamefont{E.}~\bibnamefont{Berti}},
  \bibinfo{author}{\bibfnamefont{V.}~\bibnamefont{Cardoso}}, \bibnamefont{and}
  \bibinfo{author}{\bibfnamefont{C.~M.} \bibnamefont{Will}},
  \bibinfo{journal}{Phys. Rev. D} \textbf{\bibinfo{volume}{73}},
  \bibinfo{pages}{064030} (\bibinfo{year}{2006}{\natexlab{a}}),
  \urlprefix\url{http://link.aps.org/doi/10.1103/PhysRevD.73.064030}.

\bibitem[{\citenamefont{Berti et~al.}(2007{\natexlab{a}})\citenamefont{Berti,
  Cardoso, Cardoso, and Cavaglia}}]{Berti:2007zu}
\bibinfo{author}{\bibfnamefont{E.}~\bibnamefont{Berti}},
  \bibinfo{author}{\bibfnamefont{J.}~\bibnamefont{Cardoso}},
  \bibinfo{author}{\bibfnamefont{V.}~\bibnamefont{Cardoso}}, \bibnamefont{and}
  \bibinfo{author}{\bibfnamefont{M.}~\bibnamefont{Cavaglia}},
  \bibinfo{journal}{Phys. Rev.} \textbf{\bibinfo{volume}{D76}},
  \bibinfo{pages}{104044} (\bibinfo{year}{2007}{\natexlab{a}}),
  \eprint{0707.1202}.

\bibitem[{\citenamefont{Dreyer et~al.}(2004)\citenamefont{Dreyer, Kelly,
  Krishnan, Finn, Garrison, and Lopez-Aleman}}]{Dreyer:2003bv}
\bibinfo{author}{\bibfnamefont{O.}~\bibnamefont{Dreyer}},
  \bibinfo{author}{\bibfnamefont{B.~J.} \bibnamefont{Kelly}},
  \bibinfo{author}{\bibfnamefont{B.}~\bibnamefont{Krishnan}},
  \bibinfo{author}{\bibfnamefont{L.~S.} \bibnamefont{Finn}},
  \bibinfo{author}{\bibfnamefont{D.}~\bibnamefont{Garrison}}, \bibnamefont{and}
  \bibinfo{author}{\bibfnamefont{R.}~\bibnamefont{Lopez-Aleman}},
  \bibinfo{journal}{Class. Quant. Grav.} \textbf{\bibinfo{volume}{21}},
  \bibinfo{pages}{787} (\bibinfo{year}{2004}), \eprint{gr-qc/0309007}.

\bibitem[{\citenamefont{Gossan et~al.}(2012)\citenamefont{Gossan, Veitch, and
  Sathyaprakash}}]{Gossan:2011ha}
\bibinfo{author}{\bibfnamefont{S.}~\bibnamefont{Gossan}},
  \bibinfo{author}{\bibfnamefont{J.}~\bibnamefont{Veitch}}, \bibnamefont{and}
  \bibinfo{author}{\bibfnamefont{B.~S.} \bibnamefont{Sathyaprakash}},
  \bibinfo{journal}{Phys. Rev.} \textbf{\bibinfo{volume}{D85}},
  \bibinfo{pages}{124056} (\bibinfo{year}{2012}), \eprint{1111.5819}.

\bibitem[{\citenamefont{Kamaretsos et~al.}(2012)\citenamefont{Kamaretsos,
  Hannam, Husa, and Sathyaprakash}}]{Kamaretsos:2011um}
\bibinfo{author}{\bibfnamefont{I.}~\bibnamefont{Kamaretsos}},
  \bibinfo{author}{\bibfnamefont{M.}~\bibnamefont{Hannam}},
  \bibinfo{author}{\bibfnamefont{S.}~\bibnamefont{Husa}}, \bibnamefont{and}
  \bibinfo{author}{\bibfnamefont{B.~S.} \bibnamefont{Sathyaprakash}},
  \bibinfo{journal}{Phys. Rev.} \textbf{\bibinfo{volume}{D85}},
  \bibinfo{pages}{024018} (\bibinfo{year}{2012}), \eprint{1107.0854}.

\bibitem[{\citenamefont{Nakano et~al.}(2015)\citenamefont{Nakano, Tanaka, and
  Nakamura}}]{Nakano:2015uja}
\bibinfo{author}{\bibfnamefont{H.}~\bibnamefont{Nakano}},
  \bibinfo{author}{\bibfnamefont{T.}~\bibnamefont{Tanaka}}, \bibnamefont{and}
  \bibinfo{author}{\bibfnamefont{T.}~\bibnamefont{Nakamura}},
  \bibinfo{journal}{Phys. Rev.} \textbf{\bibinfo{volume}{D92}},
  \bibinfo{pages}{064003} (\bibinfo{year}{2015}), \eprint{1506.00560}.

\bibitem[{\citenamefont{Bhagwat et~al.}(2016)\citenamefont{Bhagwat, Brown, and
  Ballmer}}]{Bhagwat:2016ntk}
\bibinfo{author}{\bibfnamefont{S.}~\bibnamefont{Bhagwat}},
  \bibinfo{author}{\bibfnamefont{D.~A.} \bibnamefont{Brown}}, \bibnamefont{and}
  \bibinfo{author}{\bibfnamefont{S.~W.} \bibnamefont{Ballmer}},
  \bibinfo{journal}{Phys. Rev.} \textbf{\bibinfo{volume}{D94}},
  \bibinfo{pages}{084024} (\bibinfo{year}{2016}), \eprint{1607.07845}.

\bibitem[{\citenamefont{Yang et~al.}(2017)\citenamefont{Yang, Yagi, Blackman,
  Lehner, Paschalidis, Pretorius, and Yunes}}]{Yang:2017zxs}
\bibinfo{author}{\bibfnamefont{H.}~\bibnamefont{Yang}},
  \bibinfo{author}{\bibfnamefont{K.}~\bibnamefont{Yagi}},
  \bibinfo{author}{\bibfnamefont{J.}~\bibnamefont{Blackman}},
  \bibinfo{author}{\bibfnamefont{L.}~\bibnamefont{Lehner}},
  \bibinfo{author}{\bibfnamefont{V.}~\bibnamefont{Paschalidis}},
  \bibinfo{author}{\bibfnamefont{F.}~\bibnamefont{Pretorius}},
  \bibnamefont{and} \bibinfo{author}{\bibfnamefont{N.}~\bibnamefont{Yunes}}
  (\bibinfo{year}{2017}), \eprint{1701.05808}.

\bibitem[{\citenamefont{Cardoso et~al.}(2016)\citenamefont{Cardoso, Franzin,
  and Pani}}]{Cardoso:2016rao}
\bibinfo{author}{\bibfnamefont{V.}~\bibnamefont{Cardoso}},
  \bibinfo{author}{\bibfnamefont{E.}~\bibnamefont{Franzin}}, \bibnamefont{and}
  \bibinfo{author}{\bibfnamefont{P.}~\bibnamefont{Pani}},
  \bibinfo{journal}{Phys. Rev. Lett.} \textbf{\bibinfo{volume}{116}},
  \bibinfo{pages}{171101} (\bibinfo{year}{2016}), \bibinfo{note}{[Erratum:
  Phys. Rev. Lett.117,no.8,089902(2016)]}, \eprint{1602.07309}.

\bibitem[{\citenamefont{Konoplya and Zhidenko}(2016)}]{Konoplya:2016hmd}
\bibinfo{author}{\bibfnamefont{R.~A.} \bibnamefont{Konoplya}} \bibnamefont{and}
  \bibinfo{author}{\bibfnamefont{A.}~\bibnamefont{Zhidenko}},
  \bibinfo{journal}{JCAP} \textbf{\bibinfo{volume}{1612}}, \bibinfo{pages}{043}
  (\bibinfo{year}{2016}), \eprint{1606.00517}.

\bibitem[{\citenamefont{Chirenti and Rezzolla}(2016)}]{Chirenti:2016hzd}
\bibinfo{author}{\bibfnamefont{C.}~\bibnamefont{Chirenti}} \bibnamefont{and}
  \bibinfo{author}{\bibfnamefont{L.}~\bibnamefont{Rezzolla}},
  \bibinfo{journal}{Phys. Rev.} \textbf{\bibinfo{volume}{D94}},
  \bibinfo{pages}{084016} (\bibinfo{year}{2016}), \eprint{1602.08759}.

\bibitem[{\citenamefont{Aasi et~al.}(2015)\citenamefont{Aasi, Abadie, Abbott,
  Abbott, Abbott, Abernathy, Accadia, Acernese, Adams, Adams
  et~al.}}]{0264-9381-32-11-115012}
\bibinfo{author}{\bibfnamefont{J.}~\bibnamefont{Aasi}},
  \bibinfo{author}{\bibfnamefont{J.}~\bibnamefont{Abadie}},
  \bibinfo{author}{\bibfnamefont{B.~P.} \bibnamefont{Abbott}},
  \bibinfo{author}{\bibfnamefont{R.}~\bibnamefont{Abbott}},
  \bibinfo{author}{\bibfnamefont{T.}~\bibnamefont{Abbott}},
  \bibinfo{author}{\bibfnamefont{M.~R.} \bibnamefont{Abernathy}},
  \bibinfo{author}{\bibfnamefont{T.}~\bibnamefont{Accadia}},
  \bibinfo{author}{\bibfnamefont{F.}~\bibnamefont{Acernese}},
  \bibinfo{author}{\bibfnamefont{C.}~\bibnamefont{Adams}},
  \bibinfo{author}{\bibfnamefont{T.}~\bibnamefont{Adams}},
  \bibnamefont{et~al.}, \bibinfo{journal}{Classical and Quantum Gravity}
  \textbf{\bibinfo{volume}{32}}, \bibinfo{pages}{115012}
  (\bibinfo{year}{2015}),
  \urlprefix\url{http://stacks.iop.org/0264-9381/32/i=11/a=115012}.

\bibitem[{\citenamefont{Acernese et~al.}(2015)\citenamefont{Acernese, Agathos,
  Agatsuma, Aisa, Allemandou, Allocca, Amarni, Astone, Balestri, Ballardin
  et~al.}}]{0264-9381-32-2-024001}
\bibinfo{author}{\bibfnamefont{F.}~\bibnamefont{Acernese}},
  \bibinfo{author}{\bibfnamefont{M.}~\bibnamefont{Agathos}},
  \bibinfo{author}{\bibfnamefont{K.}~\bibnamefont{Agatsuma}},
  \bibinfo{author}{\bibfnamefont{D.}~\bibnamefont{Aisa}},
  \bibinfo{author}{\bibfnamefont{N.}~\bibnamefont{Allemandou}},
  \bibinfo{author}{\bibfnamefont{A.}~\bibnamefont{Allocca}},
  \bibinfo{author}{\bibfnamefont{J.}~\bibnamefont{Amarni}},
  \bibinfo{author}{\bibfnamefont{P.}~\bibnamefont{Astone}},
  \bibinfo{author}{\bibfnamefont{G.}~\bibnamefont{Balestri}},
  \bibinfo{author}{\bibfnamefont{G.}~\bibnamefont{Ballardin}},
  \bibnamefont{et~al.}, \bibinfo{journal}{Classical and Quantum Gravity}
  \textbf{\bibinfo{volume}{32}}, \bibinfo{pages}{024001}
  (\bibinfo{year}{2015}),
  \urlprefix\url{http://stacks.iop.org/0264-9381/32/i=2/a=024001}.

\bibitem[{\citenamefont{Somiya}(2012)}]{0264-9381-29-12-124007}
\bibinfo{author}{\bibfnamefont{K.}~\bibnamefont{Somiya}},
  \bibinfo{journal}{Classical and Quantum Gravity}
  \textbf{\bibinfo{volume}{29}}, \bibinfo{pages}{124007}
  (\bibinfo{year}{2012}),
  \urlprefix\url{http://stacks.iop.org/0264-9381/29/i=12/a=124007}.

\bibitem[{LIG()}]{LIGOWhite}
\bibinfo{howpublished}{\url{https://dcc.ligo.org/public/0120/T1500290/002/
  T1500290.pdf}}.

\bibitem[{\citenamefont{Berti et~al.}(2009)\citenamefont{Berti, Cardoso, and
  Starinets}}]{Berti:2009kk}
\bibinfo{author}{\bibfnamefont{E.}~\bibnamefont{Berti}},
  \bibinfo{author}{\bibfnamefont{V.}~\bibnamefont{Cardoso}}, \bibnamefont{and}
  \bibinfo{author}{\bibfnamefont{A.~O.} \bibnamefont{Starinets}},
  \bibinfo{journal}{Class. Quant. Grav.} \textbf{\bibinfo{volume}{26}},
  \bibinfo{pages}{163001} (\bibinfo{year}{2009}), \eprint{0905.2975}.

\bibitem[{\citenamefont{Sathyaprakash and Schutz}(2009)}]{Sathyaprakash:2009xs}
\bibinfo{author}{\bibfnamefont{B.~S.} \bibnamefont{Sathyaprakash}}
  \bibnamefont{and} \bibinfo{author}{\bibfnamefont{B.~F.}
  \bibnamefont{Schutz}}, \bibinfo{journal}{Living Rev. Rel.}
  \textbf{\bibinfo{volume}{12}}, \bibinfo{pages}{2} (\bibinfo{year}{2009}),
  \eprint{0903.0338}.

\bibitem[{\citenamefont{Berti et~al.}(2006{\natexlab{b}})\citenamefont{Berti,
  Cardoso, and Casals}}]{Berti:2005gp}
\bibinfo{author}{\bibfnamefont{E.}~\bibnamefont{Berti}},
  \bibinfo{author}{\bibfnamefont{V.}~\bibnamefont{Cardoso}}, \bibnamefont{and}
  \bibinfo{author}{\bibfnamefont{M.}~\bibnamefont{Casals}},
  \bibinfo{journal}{Phys. Rev.} \textbf{\bibinfo{volume}{D73}},
  \bibinfo{pages}{024013} (\bibinfo{year}{2006}{\natexlab{b}}),
  \bibinfo{note}{[Erratum: Phys. Rev.D73,109902(2006)]},
  \eprint{gr-qc/0511111}.

\bibitem[{\citenamefont{Cutler and Flanagan}(1994)}]{PhysRevD.49.2658}
\bibinfo{author}{\bibfnamefont{C.}~\bibnamefont{Cutler}} \bibnamefont{and}
  \bibinfo{author}{\bibfnamefont{E.~E.} \bibnamefont{Flanagan}},
  \bibinfo{journal}{Phys. Rev. D} \textbf{\bibinfo{volume}{49}},
  \bibinfo{pages}{2658} (\bibinfo{year}{1994}),
  \urlprefix\url{http://link.aps.org/doi/10.1103/PhysRevD.49.2658}.

\bibitem[{\citenamefont{Berti et~al.}(2016)\citenamefont{Berti, Sesana,
  Barausse, Cardoso, and Belczynski}}]{Berti:2016lat}
\bibinfo{author}{\bibfnamefont{E.}~\bibnamefont{Berti}},
  \bibinfo{author}{\bibfnamefont{A.}~\bibnamefont{Sesana}},
  \bibinfo{author}{\bibfnamefont{E.}~\bibnamefont{Barausse}},
  \bibinfo{author}{\bibfnamefont{V.}~\bibnamefont{Cardoso}}, \bibnamefont{and}
  \bibinfo{author}{\bibfnamefont{K.}~\bibnamefont{Belczynski}},
  \bibinfo{journal}{Phys. Rev. Lett.} \textbf{\bibinfo{volume}{117}},
  \bibinfo{pages}{101102} (\bibinfo{year}{2016}), \eprint{1605.09286}.

\bibitem[{\citenamefont{Berti et~al.}(2007{\natexlab{b}})\citenamefont{Berti,
  Cardoso, Gonzalez, Sperhake, Hannam, Husa, and Bruegmann}}]{Berti:2007fi}
\bibinfo{author}{\bibfnamefont{E.}~\bibnamefont{Berti}},
  \bibinfo{author}{\bibfnamefont{V.}~\bibnamefont{Cardoso}},
  \bibinfo{author}{\bibfnamefont{J.~A.} \bibnamefont{Gonzalez}},
  \bibinfo{author}{\bibfnamefont{U.}~\bibnamefont{Sperhake}},
  \bibinfo{author}{\bibfnamefont{M.}~\bibnamefont{Hannam}},
  \bibinfo{author}{\bibfnamefont{S.}~\bibnamefont{Husa}}, \bibnamefont{and}
  \bibinfo{author}{\bibfnamefont{B.}~\bibnamefont{Bruegmann}},
  \bibinfo{journal}{Phys. Rev.} \textbf{\bibinfo{volume}{D76}},
  \bibinfo{pages}{064034} (\bibinfo{year}{2007}{\natexlab{b}}),
  \eprint{gr-qc/0703053}.

\bibitem[{zer()}]{zerodet}
\bibinfo{howpublished}{\url{
  https://dcc.ligo.org/cgi-bin/DocDB/ShowDocument?docid=2974}}.

\bibitem[{vir()}]{virgo}
\bibinfo{howpublished}{\url{ https://dcc.ligo.org/LIGO-P1200087-v19/public}}.

\bibitem[{kag()}]{kagra}
\bibinfo{howpublished}{\url{ http://gwcenter.icrr.u-tokyo.ac.jp/en/researcher/
  parameter}}.

\bibitem[{\citenamefont{Hild et~al.}(2010)\citenamefont{Hild, Chelkowski,
  Freise, Franc, Morgado, Flaminio, and DeSalvo}}]{Hild:2009ns}
\bibinfo{author}{\bibfnamefont{S.}~\bibnamefont{Hild}},
  \bibinfo{author}{\bibfnamefont{S.}~\bibnamefont{Chelkowski}},
  \bibinfo{author}{\bibfnamefont{A.}~\bibnamefont{Freise}},
  \bibinfo{author}{\bibfnamefont{J.}~\bibnamefont{Franc}},
  \bibinfo{author}{\bibfnamefont{N.}~\bibnamefont{Morgado}},
  \bibinfo{author}{\bibfnamefont{R.}~\bibnamefont{Flaminio}}, \bibnamefont{and}
  \bibinfo{author}{\bibfnamefont{R.}~\bibnamefont{DeSalvo}},
  \bibinfo{journal}{Class. Quant. Grav.} \textbf{\bibinfo{volume}{27}},
  \bibinfo{pages}{015003} (\bibinfo{year}{2010}), \eprint{0906.2655}.

\bibitem[{\citenamefont{Miller et~al.}(2015)\citenamefont{Miller, Barsotti,
  Vitale, Fritschel, Evans, and Sigg}}]{Miller:2014kma}
\bibinfo{author}{\bibfnamefont{J.}~\bibnamefont{Miller}},
  \bibinfo{author}{\bibfnamefont{L.}~\bibnamefont{Barsotti}},
  \bibinfo{author}{\bibfnamefont{S.}~\bibnamefont{Vitale}},
  \bibinfo{author}{\bibfnamefont{P.}~\bibnamefont{Fritschel}},
  \bibinfo{author}{\bibfnamefont{M.}~\bibnamefont{Evans}}, \bibnamefont{and}
  \bibinfo{author}{\bibfnamefont{D.}~\bibnamefont{Sigg}},
  \bibinfo{journal}{Phys. Rev.} \textbf{\bibinfo{volume}{D91}},
  \bibinfo{pages}{062005} (\bibinfo{year}{2015}), \eprint{1410.5882}.

\bibitem[{\citenamefont{Abbott et~al.}(2017)\citenamefont{Abbott, Abbott,
  Abbott, Abernathy, Ackley, Adams, Addesso, Adhikari, Adya, Affeldt
  et~al.}}]{0264-9381-34-4-044001}
\bibinfo{author}{\bibfnamefont{B.~P.} \bibnamefont{Abbott}},
  \bibinfo{author}{\bibfnamefont{R.}~\bibnamefont{Abbott}},
  \bibinfo{author}{\bibfnamefont{T.~D.} \bibnamefont{Abbott}},
  \bibinfo{author}{\bibfnamefont{M.~R.} \bibnamefont{Abernathy}},
  \bibinfo{author}{\bibfnamefont{K.}~\bibnamefont{Ackley}},
  \bibinfo{author}{\bibfnamefont{C.}~\bibnamefont{Adams}},
  \bibinfo{author}{\bibfnamefont{P.}~\bibnamefont{Addesso}},
  \bibinfo{author}{\bibfnamefont{R.~X.} \bibnamefont{Adhikari}},
  \bibinfo{author}{\bibfnamefont{V.~B.} \bibnamefont{Adya}},
  \bibinfo{author}{\bibfnamefont{C.}~\bibnamefont{Affeldt}},
  \bibnamefont{et~al.}, \bibinfo{journal}{Classical and Quantum Gravity}
  \textbf{\bibinfo{volume}{34}}, \bibinfo{pages}{044001}
  (\bibinfo{year}{2017}),
  \urlprefix\url{http://stacks.iop.org/0264-9381/34/i=4/a=044001}.

\bibitem[{\citenamefont{Abbott
  et~al.}(2016{\natexlab{c}})}]{TheLIGOScientific:2016wfe}
\bibinfo{author}{\bibfnamefont{B.~P.} \bibnamefont{Abbott}}
  \bibnamefont{et~al.} (\bibinfo{collaboration}{Virgo, LIGO Scientific}),
  \bibinfo{journal}{Phys. Rev. Lett.} \textbf{\bibinfo{volume}{116}},
  \bibinfo{pages}{241102} (\bibinfo{year}{2016}{\natexlab{c}}),
  \eprint{1602.03840}.

\bibitem[{\citenamefont{Abbott
  et~al.}(2016{\natexlab{d}})}]{TheLIGOScientific:2016src}
\bibinfo{author}{\bibfnamefont{B.~P.} \bibnamefont{Abbott}}
  \bibnamefont{et~al.} (\bibinfo{collaboration}{Virgo, LIGO Scientific}),
  \bibinfo{journal}{Phys. Rev. Lett.} \textbf{\bibinfo{volume}{116}},
  \bibinfo{pages}{221101} (\bibinfo{year}{2016}{\natexlab{d}}),
  \eprint{1602.03841}.

\bibitem[{\citenamefont{Buonanno et~al.}(2007)\citenamefont{Buonanno, Cook, and
  Pretorius}}]{PhysRevD.75.124018}
\bibinfo{author}{\bibfnamefont{A.}~\bibnamefont{Buonanno}},
  \bibinfo{author}{\bibfnamefont{G.~B.} \bibnamefont{Cook}}, \bibnamefont{and}
  \bibinfo{author}{\bibfnamefont{F.}~\bibnamefont{Pretorius}},
  \bibinfo{journal}{Phys. Rev. D} \textbf{\bibinfo{volume}{75}},
  \bibinfo{pages}{124018} (\bibinfo{year}{2007}),
  \urlprefix\url{http://link.aps.org/doi/10.1103/PhysRevD.75.124018}.

\bibitem[{\citenamefont{Berti et~al.}(2007{\natexlab{c}})\citenamefont{Berti,
  Cardoso, Gonzalez, Sperhake, Hannam, Husa, and
  Br\"ugmann}}]{PhysRevD.76.064034}
\bibinfo{author}{\bibfnamefont{E.}~\bibnamefont{Berti}},
  \bibinfo{author}{\bibfnamefont{V.}~\bibnamefont{Cardoso}},
  \bibinfo{author}{\bibfnamefont{J.~A.} \bibnamefont{Gonzalez}},
  \bibinfo{author}{\bibfnamefont{U.}~\bibnamefont{Sperhake}},
  \bibinfo{author}{\bibfnamefont{M.}~\bibnamefont{Hannam}},
  \bibinfo{author}{\bibfnamefont{S.}~\bibnamefont{Husa}}, \bibnamefont{and}
  \bibinfo{author}{\bibfnamefont{B.}~\bibnamefont{Br\"ugmann}},
  \bibinfo{journal}{Phys. Rev. D} \textbf{\bibinfo{volume}{76}},
  \bibinfo{pages}{064034} (\bibinfo{year}{2007}{\natexlab{c}}),
  \urlprefix\url{http://link.aps.org/doi/10.1103/PhysRevD.76.064034}.

\bibitem[{\citenamefont{Healy et~al.}(2013)\citenamefont{Healy, Laguna,
  Pekowsky, and Shoemaker}}]{Healy:2013jza}
\bibinfo{author}{\bibfnamefont{J.}~\bibnamefont{Healy}},
  \bibinfo{author}{\bibfnamefont{P.}~\bibnamefont{Laguna}},
  \bibinfo{author}{\bibfnamefont{L.}~\bibnamefont{Pekowsky}}, \bibnamefont{and}
  \bibinfo{author}{\bibfnamefont{D.}~\bibnamefont{Shoemaker}},
  \bibinfo{journal}{Phys. Rev.} \textbf{\bibinfo{volume}{D88}},
  \bibinfo{pages}{024034} (\bibinfo{year}{2013}), \eprint{1302.6953}.

\bibitem[{\citenamefont{Clark et~al.}(2016)\citenamefont{Clark, Bauswein,
  Stergioulas, and Shoemaker}}]{Clark:2015zxa}
\bibinfo{author}{\bibfnamefont{J.~A.} \bibnamefont{Clark}},
  \bibinfo{author}{\bibfnamefont{A.}~\bibnamefont{Bauswein}},
  \bibinfo{author}{\bibfnamefont{N.}~\bibnamefont{Stergioulas}},
  \bibnamefont{and}
  \bibinfo{author}{\bibfnamefont{D.}~\bibnamefont{Shoemaker}},
  \bibinfo{journal}{Class. Quant. Grav.} \textbf{\bibinfo{volume}{33}},
  \bibinfo{pages}{085003} (\bibinfo{year}{2016}), \eprint{1509.08522}.

\bibitem[{\citenamefont{Dominik et~al.}(2012)\citenamefont{Dominik, Belczynski,
  Fryer, Holz, Berti, Bulik, Mandel, and O'Shaughnessy}}]{Dominik:2012kk}
\bibinfo{author}{\bibfnamefont{M.}~\bibnamefont{Dominik}},
  \bibinfo{author}{\bibfnamefont{K.}~\bibnamefont{Belczynski}},
  \bibinfo{author}{\bibfnamefont{C.}~\bibnamefont{Fryer}},
  \bibinfo{author}{\bibfnamefont{D.}~\bibnamefont{Holz}},
  \bibinfo{author}{\bibfnamefont{E.}~\bibnamefont{Berti}},
  \bibinfo{author}{\bibfnamefont{T.}~\bibnamefont{Bulik}},
  \bibinfo{author}{\bibfnamefont{I.}~\bibnamefont{Mandel}}, \bibnamefont{and}
  \bibinfo{author}{\bibfnamefont{R.}~\bibnamefont{O'Shaughnessy}},
  \bibinfo{journal}{Astrophys. J.} \textbf{\bibinfo{volume}{759}},
  \bibinfo{pages}{52} (\bibinfo{year}{2012}), \eprint{1202.4901}.

\bibitem[{\citenamefont{Dominik et~al.}(2015)\citenamefont{Dominik, Berti,
  O'Shaughnessy, Mandel, Belczynski, Fryer, Holz, Bulik, and
  Pannarale}}]{Dominik:2014yma}
\bibinfo{author}{\bibfnamefont{M.}~\bibnamefont{Dominik}},
  \bibinfo{author}{\bibfnamefont{E.}~\bibnamefont{Berti}},
  \bibinfo{author}{\bibfnamefont{R.}~\bibnamefont{O'Shaughnessy}},
  \bibinfo{author}{\bibfnamefont{I.}~\bibnamefont{Mandel}},
  \bibinfo{author}{\bibfnamefont{K.}~\bibnamefont{Belczynski}},
  \bibinfo{author}{\bibfnamefont{C.}~\bibnamefont{Fryer}},
  \bibinfo{author}{\bibfnamefont{D.}~\bibnamefont{Holz}},
  \bibinfo{author}{\bibfnamefont{T.}~\bibnamefont{Bulik}}, \bibnamefont{and}
  \bibinfo{author}{\bibfnamefont{F.}~\bibnamefont{Pannarale}},
  \bibinfo{journal}{Astrophys. J.} \textbf{\bibinfo{volume}{806}},
  \bibinfo{pages}{263} (\bibinfo{year}{2015}), \eprint{1405.7016}.

\end{thebibliography}

\end{document}